\documentclass[10pt, final, conference, letterpaper, onecolumn, oneside]{./IEEEtran}

\usepackage{cite}
\usepackage{graphicx}
\usepackage{subfigure}
\usepackage{amsmath}
\usepackage{amssymb}
\usepackage{bm}
\usepackage{epsfig}
\usepackage{times}
\usepackage{color}
\usepackage{url}
\usepackage{array}
\usepackage{multirow}
\usepackage{rotating}
\usepackage{extarrows}
\usepackage{./clrscode_mod}

\newtheorem{pavikc}{\textbf{Corollary}}
\newtheorem{pavikl}{\textbf{Lemma}}

\newcommand{\argmin}{\operatornamewithlimits{argmin}}
\newcommand{\argmax}{\operatornamewithlimits{argmax}}

\begin{document}
\title{Worst-case Asymmetric Distributed Source Coding}%
\author{\IEEEauthorblockN{Samar Agnihotri\IEEEauthorrefmark{1} and Rajesh Venkatachalapathy\IEEEauthorrefmark{2}}%
\IEEEauthorblockA{\IEEEauthorrefmark{1}School of C\&EE, Indian Institute of Technology Mandi, Mandi 175001, Himachal Pradesh, India\\}
\IEEEauthorblockA{\IEEEauthorrefmark{2}Systems Science Graduate Program, Portland State University, Portland, OR 97207\\}
Email: samar.agnihotri@gmail.com, chennaipattanathar@gmail.com%
}

\maketitle

\begin{abstract}
We consider a worst-case asymmetric distributed source coding problem where an information sink communicates with $N$ correlated information sources to gather their data. A data-vector $\overline{x} = (x_1, \ldots, x_N) \sim {\mathcal P}$ is derived from a discrete and finite joint probability distribution ${\mathcal P} = p(x_1, \ldots, x_N)$ and component $x_i$ is revealed to the $i^{\textrm{th}}$ source, $1 \le i \le N$. We consider an \textit{asymmetric communication} scenario where only the sink is assumed to know distribution $\mathcal P$. We are interested in computing the minimum number of bits that the sources must send, \textit{in the worst-case}, to enable the sink to losslessly learn \textit{any} $\overline{x}$ revealed to the sources.

We propose a novel information measure called \textit{information ambiguity} to perform the worst-case information-theoretic analysis and prove its various properties. Then, we provide interactive communication protocols to solve the above problem in two different communication scenarios. We also investigate the role of block-coding in the worst-case analysis of distributed compression problem and prove that it offers \textit{almost} no compression advantage compared to the scenarios where this problem is addressed, as in this paper, with only a single instance of data-vector.

\bigskip
\end{abstract}

\section{Introduction}
\label{sec:Intro}
It is more than sixty years since Claude Shannon proposed his formulation of \textit{Information Theory} \cite{048shannon}. During intervening years, the information theory has found relevance in disciplines as diverse as Communication Theory, Theory of computation, Physics, Neural Information Processing Systems, Statistical inference and learning, and Control Theory, to name a few. Although the origins of Shannon information lie in the search for solutions to specific compression and communication problems, Shannon's information measure found itself being used in attempts to solve almost all compression and communication problems. Applications of the information theory in complex communication scenarios in diverse disciplines lead to only a few instances of successful applications \cite{098hajekEphremides}.

For the information theory to fulfill its promise as a systems notion, it must be able to produce successful results in diverse communication systems composed of heterogeneous communicating agents. Taking a cue from Shannon's solution to his original problem to understand conditions under which error free transmission of messages can take place between an information source and sink, we attempt to pose problems in communication scenarios as attempts to reach a set of design objectives the system must satisfy. In doing so, we come across useful recurring quantities that look like information measures, as in Shannon's source and channel coding theorems. 

Posing the problem of information transfer in a communication system as a question of whether a set of design specifications can be met also allows us to characterize arbitrarily complex communication scenarios in terms of very general optimization problems that include various system-level constraints, such as energy, computation and communication resources and delay-tolerance; and system characteristics, such as interaction among constituent agents and lack of global-knowledge of agents.  \textit{Real} systems often operate under such constraints and possess such characteristics. However, current information-theoretic approaches and methods often ignore such system constraints and characteristics while attempting to address more realistic models of real-world systems. This disregard for such system-level details while still attempting to use the classical (Shannon) information-theoretic results in various communication scenarios is, according to us, the primary reason for the apparent failure of the information theory in making meaningful contributions to various disciplines.

We attempt to find systematic and principled generalizations of the information theory which take into account the various resource constraints and general characteristics mentioned above. We argue that such generalizations are essential to analyze realistic models of real world compression and communication systems of diverse kinds and may result in new definitions of information, novel measures to quantify information transfer, and new variants of classical information-theoretic problems to model wider class of real systems. Recently, the need for such generalizations has been realized, resulting in some new approaches in this direction, \cite{106klir, 107flows, 108ahlswede}.

We concern ourselves with one such generalization of the classical information-theoretic problem of \textit{Distributed Source Coding (DSC)}. We first propose a new canonical scheme to classify numerous variants of the classical DSC problem. Then we consider one such variant and introduce a new information measure to aid in the analysis of this variant.

\begin{figure}[!t]
\centering
\includegraphics[width=5.25in]{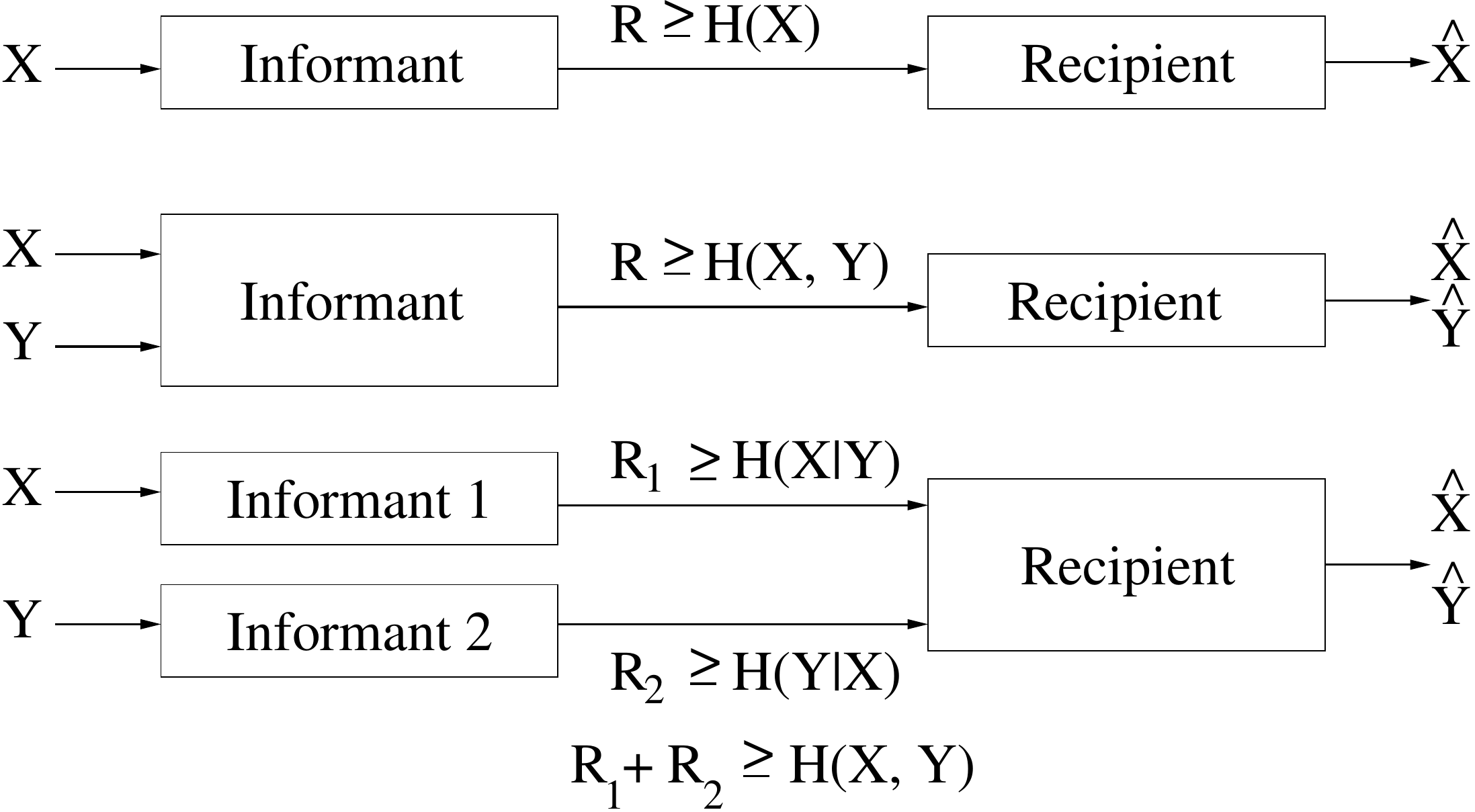}
\caption{Slepian-Wolf Distributed Source Coding problem.}
\label{fig:intro_dsc}
\end{figure}

\subsection{Distributed Source Coding}
Shannon's source coding theorem states that if an information source observes a random variable $X \sim p(x)$, then it requires to send at least $H(X)$ bits, on average, so that the information sink can losslessly recover $X$, \cite{106coverThomas}. Generalizing it for a pair random variables $(X, Y)$, it states that if an information source observes a pair of random variables $(X, Y) \sim {\cal P} = p(x, y)$ and jointly encodes those, then $H(X, Y)$ bits from the information source, on average, are sufficient for the information sink to losslessly recover $(X, Y)$, as in Figure~\ref{fig:intro_dsc}.

However, the authors in \cite{073slepianWolf} proved a surprising result that states if two correlated random variables $(X, Y) \sim {\cal P} = p(x, y)$ are observed by two non-cooperating information sources, and $X$ and $Y$ are independently encoded, then as long as the sources send a total of $H(X, Y)$ bits, on average, it is still sufficient for the sink to losslessly recover $(X, Y)$, with individual information rates for $X$ and $Y$ being at least $H(X|Y)$ and $H(Y|X)$, respectively, as in Figure~\ref{fig:intro_dsc}.

Fundamentally, the information transfer in \textit{any} communication system where a sink node is interested in collecting information from a set of correlated information sources, which do not communicate among themselves, can be modeled as distributed source coding problem. Communication system in this context can be a model of communication in neural information processing system, learning and estimation system, or wireless communication system. However, the set of constraints and characteristics of the particular communication system and objectives of communication in the system often determine the corresponding variant of distributed source coding problem that is most appropriate to model the system in question. The resultant variants differ from each other not only in terms of the problem definition, but also in terms of the computation and communication complexities of their optimal solutions.

Since the publication of the seminal paper of Slepian and Wolf, various attempts have been made towards solving distributed source coding problem, such as \cite{104chouPetrovic, 104xiongLiveris, 092orlitsky, 105adler} and the references therein. However, there is no single definition of DSC problem. In the absence of any unified framework to systematically generate and address the different variants of distributed source coding problem, it is often difficult to compare and reconcile the approaches and solutions of different variants. We propose a canonical framework to construct and address such different variants. The proposed framework classifies each DSC problem variant according to subset of assumptions and objectives used to define the problem variant. The sets of particular assumptions and objectives we consider are as follows.

\textbf{Assumptions:}
\begin{itemize}
\item Symmetric or asymmetric communication corresponding to presence or absence of global knowledge at the information sources, respectively.
\item Interactive (with limited or unlimited number of messages) or non-interactive communication between the sink and the sources.
\item Serial or parallel communication from the sources to the sink.
\item Block-encoding of data-samples at the sources.
\end{itemize}

\textbf{Objectives:}
\begin{itemize}
\item Lossy or lossless data-gathering at the sink.
\item Worst-case or average-case performance analysis.
\item Function of number of bits communicated by a subset of communicating agents (the sink and/or the sources) that is to be minimized. Examples of such functions frequently used in defining DSC problem variants are \textit{sum} and \textit{max}.
\item The subset of communicating agents over which such minimizations is carried out.
\end{itemize}

\begin{figure}[!t]
\centering
\includegraphics[width=5.25in]{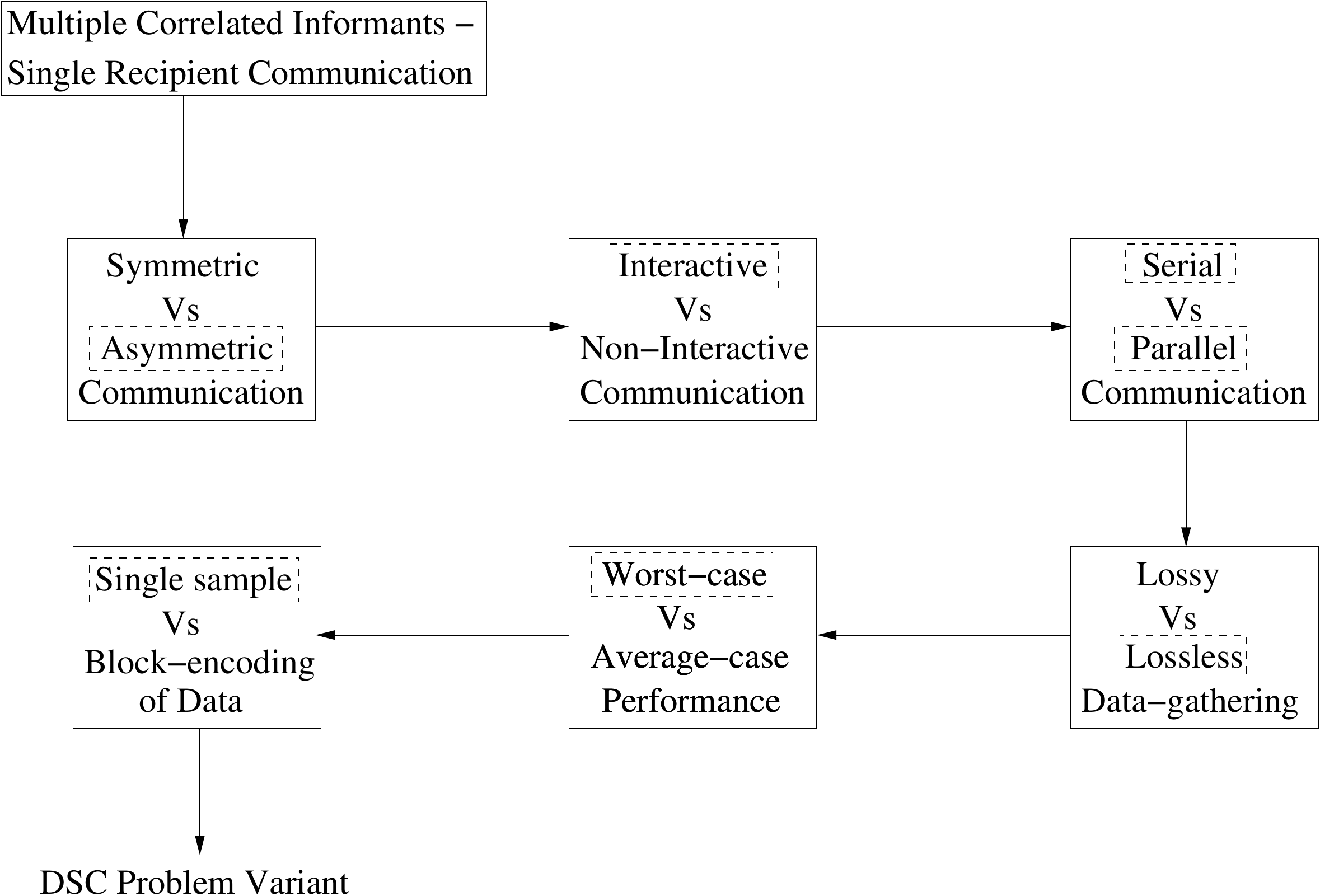}
\caption{A canonical scheme to construct variants of Distributed Source Coding problem. The variant of DSC problem considered in this paper is constructed by selecting the dashed boxes in each block.}
\label{fig:dsc_variants}
\end{figure}

This is explained pictorially in Figure~\ref{fig:dsc_variants}. For example, the variant of DSC problem considered by Slepian and Wolf in \cite{073slepianWolf} is defined by the assumptions of symmetric communication and block-coding with asymptotically large block-lengths and has the objective of lossless data-gathering at sink, average-case performance analysis, minimizing the sum of source bits. The variant of DSC problem considered in this paper is constructed by selecting the dashed boxes in each block.

It should be noted that we \textit{do not} suggest that ours is the only way to classify the variants of distributed source coding problem, as one can come-up with alternative classification schemes with fewer or more parameters, such as the canonical classification scheme proposed in \cite{108janaBlahut} that is contained in our scheme. However, we \textit{do} suggest that after almost forty years since seminal Slepian-Wolf paper, it is the time when we should systematically address DSC problem by following a unified scheme to classify its variants and construct practical solutions achieving optimal performance. To the best of our knowledge, ours is the first such scheme.

\subsection{Motivation}
Our motivation to consider distributed source coding problem, in particular a variant of it, comes from its strong connection to the problem of maximizing the worst-case operational lifetime of data-gathering wireless sensor networks \cite{102akyildiz}, where the base-station collects correlated sensor data. Sensor nodes constituting such networks are assumed to have limited battery resources that cannot be replenished either due to sensor nodes being deployed in inaccessible locations or due to high cost of retrieving sensor nodes and changing their batteries. This makes it impossible to replace the dead nodes. However such networks, once deployed, are expected to have large operational lifetimes. In such a scenario, a key challenge is to develop system-level strategies to efficiently utilize finite energy resources to prolong network lifetime.

Sensor nodes expend energy in sensing/actuating, computation, and communication. However, the harsh nature of wireless links determines the energy cost of communication that has the potential to be a major bottleneck. One of the significant factors that determines the communication energy expenditure at nodes is the number of bits exchanged in the network for successful data-gathering. Therefore any scheme that reduces the number of such bits can make significant contribution to enhance the network lifetime.

In a typical data-gathering sensor network, there are two fundamental asymmetries: \textit{resource asymmetry} and \textit{information asymmetry}. In such networks, it is reasonable to assume that the base-station has large energy, computation, and communication resources, whereas sensor nodes are resource limited (\textit{resource asymmetry}). Further, in sensor networks, the sensor nodes hold the actual sampled information and the base-station may only know the general characteristics such as joint probability distribution of sensor data (\textit{information asymmetry}). We propose that the resource asymmetry in wireless sensor networks should be exploited to reduce the information asymmetry in such networks. Therefore we argue that more resourceful and knowledgeable base-station that wants to gather sensor-data, should bear most of the burden of computation and communication in the network. Allowing interactive communication between the base-station and sensor nodes enables us to do this: base-station forms and communicates \textit{efficient} queries to the sensor nodes, which they respond to with short and easily computable messages.

Given the correlated nature of sensor data, the data-gathering problem in wireless sensor networks can be modeled in terms of well-known \textit{Distributed Source Coding (DSC)} problem \cite{073slepianWolf} or its variants. In the recent past, there have been several such attempts, such as \cite{104chouPetrovic, 105adler} and various other schemes surveyed in \cite{104xiongLiveris}. However, most of these schemes are not \textit{really} effective in the data-gathering wireless sensor network due to various reasons. First, in some scenarios we are interested in minimizing the worst-case number of sensor bits (or equivalently in maximizing the worst-case network lifetime). However, the schemes based on average-case information-theoretic analysis cannot be used in such scenarios. Second, such schemes cannot address various operational constraints and requirements of sensor network, such as non-availability of global knowledge of the system, such as joint distribution of sensor data, at the sensor nodes and low-latency operation. Third, these schemes do not attempt to exploit various opportunities, such as resource asymmetry, in such networks to reduce the computation and communication burden at nodes. Therefore to address various aforementioned shortcomings of existing distributed source coding schemes in the context of data-gathering wireless sensor networks, we introduce a new variant of the classical distributed source coding problem as follows. The proposed variant attempts to optimally utilize the resource asymmetry in a data-gathering wireless sensor network to minimize the information asymmetry in the network.

Consider a distributed information-gathering scenario, where a sink collects the information from $N$ correlated information sources. The correlation in the sources' data is modeled by joint distribution $\mathcal P$, which is known only to the sink. The sink and sources can interactively communicate with each other with communication proceeding in rounds. We are primarily concerned with minimizing the number of bits that the sources send, in the worst-case, for successful data-gathering at the sink, but we are also interested in minimizing both the number of communication rounds and the number of sink bits.

Our work mainly differs from the previous work on distributed source coding and its applications to sensor networks as follows. Firstly, we assume asymmetric communication where only the sink knows the correlation structure of sources' data. This is in contrast to existing DSC schemes that assume that all nodes know the correlation structure. Secondly, unlike existing DSC schemes that perform average-case information-theoretic analysis, we are concerned with the worst-case performance analysis of distributed source coding. As the average-case information measure of entropy or its variants cannot be used for the worst-case information-theoretic analysis, we introduce \textit{information ambiguity} - a new information measure for worst-case information-theoretic analysis. Thirdly, we are interested in distributed compression when only a single instance of data is available at every information source (\textit{oneshot compression}) unlike majority of current DSC schemes that derive their results in the regime of infinite block-lengths. Finally, we consider a more powerful model of communication where the sink and sources interactively communicate with each other. 

\textit{Note on the terminology:} We consider communication system consisting of communicating agents of two types: information sources and information sink. We address information sources also as informants and source nodes. Similarly, we also address the information sink interchangeably as the receiver and the recipient.

\subsection{Organization}
The paper is organized as follows. In Section~\ref{sec:relatedWork}, we survey the related work. Section~\ref{sec:iAmbiguity} introduces the notion of \textit{information ambiguity} for the worst-case information-theoretic analyses, discusses some of its properties, and proves that it is a valid information measure. In Section~\ref{sec:probSetting}, we provide precise description of the communication model we assume and formally introduce the distributed data-gathering problem we address in this paper. Then, Section~\ref{sec:bCompressibility4cmprsn} provides the solutions of this problem under two different communication scenarios. We first present an interactive communication protocol to optimally minimize the number of informant bits required in the worst-case to solve the problem. Later, we provide an optimal interactive parallel communication protocol that efficiently trades-off the number of informant bits to reduce the number of communication rounds and the number of sink bits. Section~\ref{sec:blkCoding4cmprsn} investigates the role of block-coding in the worst-case analysis of distributed source coding problem and proves that unlike the average-case performance, worst-case performance of DSC problems derive \textit{almost} no advantage from the block-coding compared to \textit{oneshot compression}. Finally, we conclude and discuss some future work in Section~\ref{sec:conclusions}.

\section{Related work}
\label{sec:relatedWork}
The Slepian-Wolf solution \cite{073slepianWolf} of distributed source coding problem, though fundamental, is essentially existential and non-constructive, like many other results of the classical information theory. Though it establishes the lower bounds on the information rates, it does not provide us the optimal source codes or any computationally efficient method of constructing those. Therefore, in the recent past, numerous attempts have been made to provide practical solutions for it. In \cite{103pradhanRamchandran}, the authors came up with the DISCUS framework to give practical, though not necessarily optimal, method to construct source codes. Though the result for the duality between Slepian-Wolf encoding and multiple-access channel was already well-known \cite{106coverThomas}, the connection that this piece of work made between distributed source coding and channel coding, motivated the researchers to use various channel codes, such as Turbo codes \cite{101garciaZhao, 101bajcsyMitran, 102aaronGirod}, LDPC codes \cite{102liverisXiong, 103garciaZhong, 104colemanLee, 104lanLiveris}, and Convolution codes \cite{103liverisXiong}, to solve the distributed source coding problem. In \cite{103zhaoEffros} and related papers, Zhao and Effros have addressed the lossless and near-lossless source code design and construction problem. Also, given the asymmetry in the available energy and computational resources between the base-station and the sensor nodes, \cite{105zhongRabaey} argues to use such asymmetric channel codes to reduce the energy consumption at the sensor nodes. A survey in \cite{104xiongLiveris} and the references therein provide more details about some of these research efforts.

These developments, though pragmatic and constructive, are not very practical in the context of sensor networks, particularly due to their assumption of symmetric communication scenarios, where all nodes in the network are assumed to know the joint probability distribution of sensor-data, and requirement of large coding dimensions, where a large number of independent and identically distributed (i.i.d) samples are drawn and each informant encodes the sequence of these samples as a single codeword to achieve the optimal performance. Given the limited communication and computation capabilities of the sensor nodes, it is neither reasonable to assume that the sensor nodes know the joint distribution of all sensors' data, nor to assume that sensor nodes can carry out high-complexity encoding. Also, the block-encoding with very large block-lengths (typically, $\sim \! 10^4$ data samples) required by these schemes may incur large data-gathering delays, rendering these solutions inefficient, given the time-criticality of sensor-data.

The notion of interactive communication in addressing distributed source coding problem was introduced in \cite{092orlitsky}. Later in \cite{104chouPetrovic}, authors attempted to deploy sink based feedback to construct practical schemes to address data-gathering problem in the data-gathering wireless sensor networks. However, with just a single feedback message, this work could not make use of the full potential of interactive communication in realizing optimum DSC performance in the sensor networks. Adler in \cite{105adler} is concerned with analysing the performance of distributed source coding problem in a scenario where there is only a single instance of informant-data and the sink and informants communicate interactively, however like all previous work on distributed source coding, this work also performs only average-case analysis.

The notion of information ambiguity that we propose as the worst-case equivalent of the notion of information entropy, was introduced by Orlitsky in \cite{090orlitsky}, but in a different context than ours. Also, the researchers in the field of ``Possibility Theory'' have endeavored to define some information measures, which are closely related to the notion of information ambiguity. However, it is beyond the scope of this paper to discuss those efforts and an interested reader can find the broad survey of such work in \cite{097deCooman, 106klir}.

\section{Information Ambiguity}
\label{sec:iAmbiguity}
The original Shannon's theorems and all subsequent theorems in information theory are all asymptotic results based on the Large Deviations Theory. It implies the need to have very large set of data samples and leads to what we call average-case results. Worst-case analysis deals with sparse data gathering situations and is the sole focus of this work. Leaving the precise implementation motivations for its definition for the later sections, here we define a new information measure which we call \textit{Information Ambiguity}, show that it is a valid information measure, and characterize some of its properties useful for our later results.

We begin by introducing the notion of information ambiguity for two random variables and then provide its exposition for arbitrary number of variables. Note that throughout the paper all the logarithms are to base two.

\subsection{Ambiguity: Two Random Variables}
\label{subsec:2ambiguity}
Consider a pair of random variables $(X_1, X_2) \sim {\mathcal P} = p(x_1, x_2), X_1 \in {\mathcal X}$ and $X_2 \in {\mathcal X}$, where ${\mathcal X}$ is discrete and finite alphabet of size\footnote{In general, $X_1 \in {\mathcal X}_1$ and $X_2 \in {\mathcal X}_2$, where ${\mathcal X}_1$ and ${\mathcal X}_1$ are discrete alphabet sets, with possibly different cardinalities. However, to keep the discussion simple, we assume henceforth that all the random variables take the values from the same discrete alphabet ${\mathcal X}$.} ${|\mathcal X|}$ and $\mathcal P$ is the joint probability distribution of $(X_1, X_2)$. The \textit{support set} of $(X_1, X_2)$ is defined as:
\begin{equation}
\label{eqn:supp_set_dstrbn2}
S_{X_1, X_2} \stackrel{\textrm{def}}{=} \{(x_1, x_2) | p(x_1, x_2) > 0 \}
\end{equation}

We also call $S_{X_1, X_2}$ as the \textit{ambiguity set} of $(X_1, X_2)$. The cardinality of $S_{X_1, X_2}$ is called \textit{joint ambiguity} or simply \textit{ambiguity} of $(X_1, X_2)$ and denoted as $\mu_{X_1, X_2} = |S_{X_1, X_2}|$. The minimum number of bits required to describe all elements in $S_{X_1, X_2}$ is $\lceil \log \mu_{X_1, X_2} \rceil$.

The \textit{support set} of $X_1$, is set
\begin{equation}
\label{eqn:supp_set_o1rv4dstrbn2}
S_{X_1} \stackrel{\textrm{def}}{=} \{x_1: \mbox{ for some } x_2, (x_1, x_2) \in S_{X_1, X_2}\},
\end{equation}
of all possible $X_1$ values. We also call $S_{X_1}$ \textit{ambiguity set} of $X_1$. The \textit{ambiguity} of $X_1$ is defined as $\mu_{X_1} = |S_{X_1}|$. The ambiguity set and the corresponding ambiguity of random variable $X_2$ is similarly defined.

The \textit{conditional ambiguity set} of $X_1$, when random variable $X_2$ takes the value $x_2 \in S_{X_2}$ is
\begin{equation}
\label{eqn:ambiguity_set}
S_{X_1|X_2}(x_2) \stackrel{\textrm{def}}{=} \{x_1: (x_1, x_2) \in S_{X_1, X_2} \},
\end{equation}
the set of possible $X_1$ values when $X_2 = x_2$. The \textit{conditional ambiguity} in that case is
\begin{equation}
\label{eqn:ambiguity}
\mu_{X_1|X_2}(x_2) \stackrel{\textrm{def}}{=} |S_{X_1|X_2}(x_2)|,
\end{equation}
the number of possible $X_1$ values when $X_2=x_2$. The \textit{maximum conditional ambiguity} of $X_1$ is
\begin{equation}
\label{eqn:max_ambiguity}
\widehat{\mu}_{X_1|X_2} \stackrel{\textrm{def}}{=} \max \{\mu_{X_1|X_2}(x_2): x_2 \in S_{X_2}\},
\end{equation}
the maximum number of $X_1$ values possible with any value that $X_2$ can take. We denote the corresponding \textit{maximum conditional ambiguity set} as $S_{X_1|X_2}$.

The quantities $S_{X_2|X_1}(x_1), \mu_{X_2|X_1}(x_1), S_{X_2|X_1}$, and $\widehat{\mu}_{X_2|X_1}$ are similarly defined by exchanging the roles of $X_1$ and $X_2$ in the preceding discussion.

Define a functional called \textit{information ambiguity} as ${\cal I}_{X_1, X_2} = \lceil \log \mu_{X_1, X_2} \rceil$ for a set of two random variables $X_1$ and $X_2$. Next, we prove certain properties of functional ${\cal I}_{X_1, X_2}$.

\begin{pavikl}
\label{lemma:conditionalAmbiguity}
${\cal I}_{X_1|X_2}(X_2 = x_2) \le {\cal I}_{X_1}$ for all $x_2 \in S_{X_2}$, that is, conditioning reduces information ambiguity.
\end{pavikl}
\begin{IEEEproof}
From the definitions of $S_{X_1}$ and $S_{X_1|X_2}(x_2)$, it is obvious that $S_{X_1|X_2}(x_2) \subseteq S_{X_1}$. This implies that $\mu_{X_1|X_2}(x_2) \le \mu_{X_1}$.
\end{IEEEproof}

Also, it follows from Lemma~\ref{lemma:conditionalAmbiguity} and \eqref{eqn:max_ambiguity} that $\widehat{\cal I}_{X_1|X_2} \le {\cal I}_{X_1}$.

\begin{pavikl}
\label{lemma:2subadditivity}
\textbf{(Subadditivity)} If $X_1$ and $X_2$ are ``interacting'', that is $S_{X_1, X_2} \subseteq S_{X_1} \times S_{X_2}$, then ${\cal I}_{X_1, X_2} \le {\cal I}_{X_1} + {\cal I}_{X_2}$.
\end{pavikl}
\begin{IEEEproof}
We know that $S_{X_1, X_2} \subseteq S_{X_1} \times S_{X_2}$. So,
\begin{align*}
\mu_{X_1, X_2} & \le \mu_{X_1} \times \mu_{X_2} \\
\log \mu_{X_1, X_2} & \le \log (\mu_{X_1} \times \mu_{X_2}) \\
                     & = \log \mu_{X_1} + \log \mu_{X_2} \\
\lceil \log \mu_{X_1, X_2} \rceil & \le \lceil \log \mu_{X_1} + \log \mu_{X_2} \rceil \\
                                  & \le \lceil \log \mu_{X_1} \rceil + \lceil \log \mu_{X_2} \rceil,
\end{align*}
thus, proving the lemma.
\end{IEEEproof}

\begin{pavikl}
\label{lemma:2additivity}
\textbf{(Additivity)} If $X_1$ and $X_2$ are ``non-interacting'', that is $S_{X_1, X_2} = S_{X_1} \times S_{X_2}$ then ${\cal I}_{X_1, X_2} \stackrel{\cdot}{=} {\cal I}_{X_1} + {\cal I}_{X_2}$, where $\stackrel{\cdot}{=}$ denotes equality within one bit per random variable.
\end{pavikl}
\begin{IEEEproof}
We know that $S_{X_1, X_2} = S_{X_1} \times S_{X_2}$. So,
\begin{align*}
\mu_{X_1, X_2} & = \mu_{X_1} \times \mu_{X_2} \\
\log \mu_{X_1, X_2} & = \log (\mu_{X_1} \times \mu_{X_2}) \\
                     & = \log \mu_{X_1} + \log \mu_{X_2} \\
\lceil \log \mu_{X_1, X_2} \rceil & = \lceil \log \mu_{X_1} + \log \mu_{X_2} \rceil \\
                                  & \stackrel{\cdot}{=} \lceil \log \mu_{X_1} \rceil + \lceil \log \mu_{X_2} \rceil
\end{align*}
Thus proving the lemma.
\end{IEEEproof}

\begin{pavikl}
\label{lemma:2chainRule}
${\cal I}_{X_1, X_2} \le {\cal I}_{X_1} + \widehat{\cal I}_{X_2|X_1}$
\end{pavikl}
\begin{IEEEproof}
We know that $S_{X_1, X_2} \subseteq S_{X_1} \times S_{X_2|X_1}$. So,
\begin{align*}
\mu_{X_1, X_2} & \le \mu_{X_1} \times \widehat{\mu}_{X_2|X_1} \\
\log \mu_{X_1, X_2} & \le \log \mu_{X_1} + \log \widehat{\mu}_{X_2|X_1} \\
\lceil \log \mu_{X_1, X_2} \rceil & \le \lceil \log \mu_{X_1} + \log \widehat{\mu}_{X_2|X_1} \rceil \\
                                  & \le \lceil \log \mu_{X_1} \rceil + \lceil \log \widehat{\mu}_{X_2|X_1} \rceil
\end{align*}
This proves the lemma.
\end{IEEEproof}

\begin{pavikc}
\label{corrl:2chainRule}
${\cal I}_{X_1, X_2} \le {\cal I}_{X_2} + \widehat{\cal I}_{X_1|X_2}$
\end{pavikc}
\begin{IEEEproof}
Reversing the roles of $X_1$ and $X_2$ in the proof of Lemma~\ref{lemma:2chainRule}, completes the proof.
\end{IEEEproof}

\begin{pavikl}
\label{lemma:2ambiguityBound}
Let $\Pi$ denote the set of two possible permutations of $\{1, 2\}$, then
\begin{equation*}
{\cal I}_{X_1, X_2} \le \min_{\pi \in \Pi} ({\cal I}_{X_{\pi(1)}} + \widehat{\cal I}_{X_{\pi(2)}|X_{\pi(1)}})
\end{equation*}
\end{pavikl}
\begin{IEEEproof}
The proof follows from combining Lemma~\ref{lemma:2chainRule} and Corollary~\ref{corrl:2chainRule}.
\end{IEEEproof}

\subsection{Ambiguity Computation: An Example}
\label{subsec:ambiguityExa}
\begin{figure}[!t]
\centering
\includegraphics[width=3.5in]{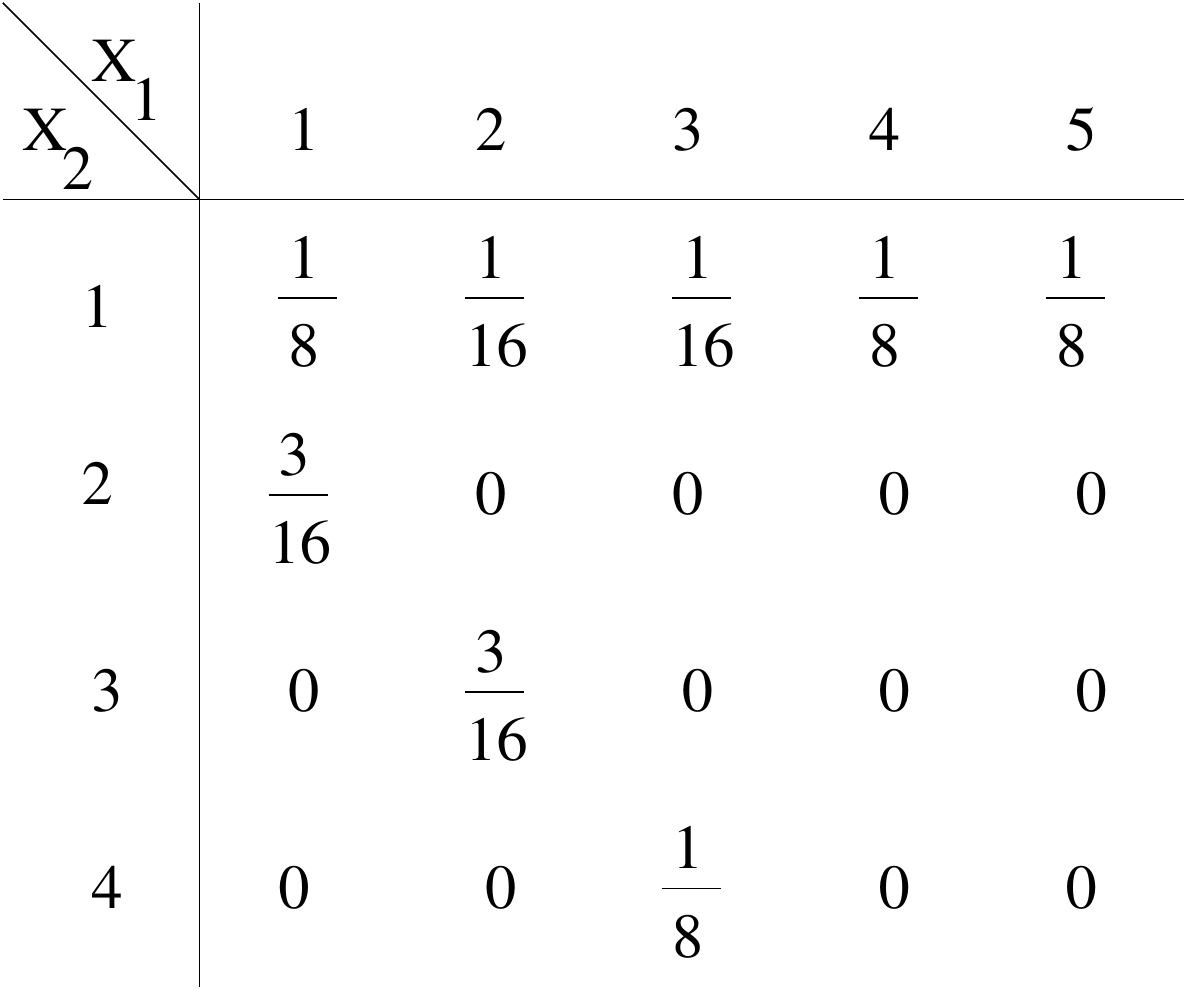}
\caption{The probability distribution $\mathcal P$ for $(X_1, X_2)$.}
\label{fig:supportSet}
\end{figure}

We illustrate some of the definitions and properties of the notion of information ambiguity we have discussed in this section, using the probability distribution $\mathcal P$ for two random variables $X_1, X_2$, given in Figure~\ref{fig:supportSet}.

Using \eqref{eqn:supp_set_dstrbn2}, support set $S_{X_1, X_2}$ is:
\begin{equation*}
S_{X_1, X_2} = \{(1,1), (1,2), (1,3), (1,4), (1,5), (2,1), (3,2), (4,3) \}
\end{equation*}
with corresponding ambiguity $\mu_{X_1, X_2} = |S_{X_1, X_2}| = 8$. Therefore, the number of bits required to describe the elements of $S_{X_1, X_2}$ are ${\cal I}_{X_1, X_2} = \lceil \log \mu_{X_1, X_2} \rceil = 3$ bits.

Further, using \eqref{eqn:supp_set_o1rv4dstrbn2} the support sets of $X_1$ and $X_2$ are, respectively:
\begin{align*}
S_{X_1} & = \{1, 2, 3, 4, 5 \}, \mu_{X_1} = |S_{X_1}| = 5 \\
S_{X_2} & = \{1, 2, 3, 4 \}, \quad \mu_{X_2} = |S_{X_2}| = 4
\end{align*}
The minimum number of bits required to describe the elements of $S_{X_1}$ and $S_{X_2}$ are ${\cal I}_{X_1} = \lceil \log \mu_{X_1} \rceil = 3$ bits and ${\cal I}_{X_2} = \lceil \log \mu_{X_2} \rceil = 2$ bits, respectively.

Computing the conditional ambiguity sets and corresponding conditional ambiguities for $X_1$, we have using \eqref{eqn:ambiguity_set}:
\begin{align*}
S_{X_1|X_2}(X_2 = 1) = \{1, 2, 3, 4, 5 \} & , \mu_{X_1|X_2}(X_2 = 1) = 5 \\
S_{X_1|X_2}(X_2 = 2) = \{1\} & , \mu_{X_1|X_2}(X_2 = 2) = 1 \\
S_{X_1|X_2}(X_2 = 3) = \{2\} & , \mu_{X_1|X_2}(X_2 = 3) = 1 \\
S_{X_1|X_2}(X_2 = 4) = \{3\} & , \mu_{X_1|X_2}(X_2 = 4) = 1
\end{align*}
Therefore, the maximum conditional ambiguity in $X_1$ given $X_2$ is $\widehat{\mu}_{{X_1|X_2}} = 5$.

Similarly, for $X_2$, we have:
\begin{align*}
S_{X_2|X_1}(X_1 = 1) = \{1, 2\} & , \mu_{X_2|X_1}(X_1 = 1) = 2 \\
S_{X_2|X_1}(X_1 = 2) = \{1, 3\} & , \mu_{X_2|X_1}(X_1 = 2) = 2 \\
S_{X_2|X_1}(X_1 = 3) = \{1, 4\} & , \mu_{X_2|X_1}(X_1 = 3) = 2 \\
S_{X_2|X_1}(X_1 = 4) = \{1\}    & , \mu_{X_2|X_1}(X_1 = 4) = 1 \\
S_{X_2|X_1}(X_1 = 5) = \{1\}    & , \mu_{X_2|X_1}(X_1 = 5) = 1
\end{align*}
Hence, the maximum conditional ambiguity in $X_2$ given $X_1$ is $\widehat{\mu}_{{X_2|X_1}} = 2$.

Above, we computed ${\cal I}_{X_1, X_2} = 3$ bits. Now, let us compute
\begin{align*}
{\cal I}_{X_1} + \widehat{\cal I}_{X_2|X_1} = 3 \mbox{ bits } + 1 \mbox{ bits } = 4 \mbox{ bits } \\
{\cal I}_{X_2} + \widehat{\cal I}_{X_1|X_2} = 2 \mbox{ bits } + 3 \mbox{ bits } = 5 \mbox{ bits }
\end{align*}
This illustrates Lemma~\ref{lemma:2chainRule}, Corollary~\ref{corrl:2chainRule}, and Lemma~\ref{lemma:2ambiguityBound}.

\subsection{Ambiguity: $N$ Random Variables}
\label{subsec:Nambiguity}
The definitions and results of subsection~\ref{subsec:2ambiguity} for the information ambiguity of two random variables are easily extended to their multiple random variable counterparts for the information ambiguity of a set of $N$ random variables $U = \{X_1, \ldots$, $X_N\}$.

Consider a discrete and finite probability distribution $\mathcal P$ for $N$ random variables $X_i, i \in {\mathcal N}, X_i \in {\mathcal X}$, where ${\mathcal N} = \{1, \ldots, N\}$ and $\mathcal X$ is the discrete and finite alphabet of size $|{\mathcal X}|$. Consider a $N$-tuple of random variables $(X_1, \ldots, X_N) \sim {\mathcal P} = p(x_1, \ldots, x_N)$. The \textit{support set} of $(X_1, \ldots, X_N)$ is defined as:
\begin{equation}
\label{eqn:supp_set_dstrbnN}
S_{X_1, \ldots, X_N} \stackrel{\textrm{def}}{=} \{(x_1, \ldots, x_N) | p(x_1, \ldots, x_N) > 0 \}
\end{equation}
We also call $S_{X_1, \ldots, X_N}$ the \textit{ambiguity set} of $(X_1, \ldots, X_N)$. The cardinality of $S_{X_1, \ldots, X_N}$ is called \textit{ambiguity} of $(X_1, \ldots, X_N)$ and denoted as $\mu_{X_1, \ldots, X_N} = |S_{X_1, \ldots, X_N}|$. Therefore the minimum number of bits required to describe an element in $S_{X_1, \ldots, X_N}$ is $\lceil \log \mu_{X_1, \ldots, X_N} \rceil$.

Consider sets of random variables $X_A$ and $X_A^c$ such that $X_A^c = U \setminus X_A$. Denote instances of $X_A$ and $X_A^c$ as $x_A$ and $x_A^c$, respectively. The support-set of $X_A$ is defined as:
\begin{equation*}
S_{X_A} \stackrel{\textrm{def}}{=} \{x_A | \mbox{ for some } x_A^c, (x_A, x_A^c) \in S_{X_1,\ldots, X_N} \}
\end{equation*}
We also call $S_{X_A}$ as the \textit{ambiguity set} of $X_A$, with corresponding \textit{ambiguity} denoted as $\mu_{X_A}$ and defined as $\mu_{X_A} = |S_{X_A}|$. So, the minimum number of bits required to describe any value of $X_A$ is $\lceil \log \mu_{X_A} \rceil$.

Consider random variable $X_i \in X_A$ and the set of random variables $X_B \subset X_A, X_i \not\in X_B$. Denote an instance of $X_B$ as $x_B$. The \textit{conditional ambiguity set} of $X_i$, when set $X_B$ takes value $x_B \in S_{X_B}$ is
\begin{equation}
\label{eqn:ambiguity_setN}
S_{X_i|X_B}(x_B) \stackrel{\textrm{def}}{=} \{x_i: (x_i, x_B) \in S_{X_A} \},
\end{equation}
the set of possible $X_i$ values when $X_B = x_B$. The \textit{conditional ambiguity} in that case is
\begin{equation}
\label{eqn:ambiguityN}
\mu_{X_i|X_B}(x_B) \stackrel{\textrm{def}}{=} |S_{X_i|X_B}(x_B)|,
\end{equation}
the number of possible $X_i$ values with $X_B = x_B$. The \textit{maximum conditional ambiguity} of $X_i$ is
\begin{equation}
\label{eqn:max_ambiguityN}
\widehat{\mu}_{X_i|X_B} \stackrel{\textrm{def}}{=} \sup \{\mu_{X_i|X_B}(x_B): x_B \in S_{X_B}\},
\end{equation}
the maximum number of $X_i$ values possible over any $x_B$.

In fact, for any two subsets $X_A$ and $X_B$ of $\{X_1, \ldots, X_N\}$, such that $X_A \cup X_B \subseteq \{X_1, \ldots, X_N\}$ and $X_A \cap X_B = \phi$, we can define for example, \textit{ambiguity set} $S_{X_A}$ of $X_A$, \textit{conditional ambiguity set} $S_{X_A|X_B}(x_B)$ of $X_A$ given the set $x_B$ of values that $X_B$ can take, and \textit{maximum conditional ambiguity set} $S_{X_A|X_B}$ of $X_A$ for any set of values that $X_B$ can take, with corresponding \textit{ambiguity}, \textit{conditional ambiguity}, and \textit{maximum conditional ambiguity} denoted by $\mu_{X_A}$, $\mu_{X_A|X_B}(x_B)$, and $\widehat{\mu}_{X_A|X_B}$, respectively. However, for the sake of brevity, we do not introduce these definitions here as those can be easily developed along the lines of the definitions in \eqref{eqn:ambiguity_setN}-\eqref{eqn:max_ambiguityN}.

Further, let us represent each of $\mu_{X_i}$ values that random variable $X_i$ can assume in $\lceil \log \mu_{X_i} \rceil$ bits as $b^i_1 \ldots b^i_{\lceil \log \mu_{X_i} \rceil}$. Let $\mbox{binary}_j(x_i)$ represent the value of the $j^{\textrm{th}}$ bit-location in the bit-representation of $x_i$, $1 \le j \le \lceil \log \mu_{X_i} \rceil$. Then, knowing that the value of $j^{\textrm{th}}$ bit-location is $b$, we can define the set of possible values that $X_i$ can take as
\begin{equation}
\label{eqn:cond_ambiguity4Bit}
S_{X_i|b^i_j}(b) \stackrel{\textrm{def}}{=} \{x_i: x_i \in S_{X_i} \mbox{ and } \mbox{binary}_j(x_i) = b\},
\end{equation}
with corresponding cardinality denoted as $\mu_{X_i|b^i_j}(b)$. We can similarly define $S_{X_A|b^i_j}(b)$ with $X_i \in X_A$ as
\begin{equation}
\label{eqn:cond_ambiguityOset4Bit}
S_{X_A|b^i_j}(b) \stackrel{\textrm{def}}{=} \{x_A: x_A \in S_{X_A} \mbox{ and } \mbox{binary}_j(x_i) = b\},
\end{equation}
with corresponding cardinality denoted as $\mu_{X_A|b^i_j}(b)$. The definitions in \eqref{eqn:cond_ambiguity4Bit} and \eqref{eqn:cond_ambiguityOset4Bit} can be easily extended further to the situations where the values of one or more bit-locations in one or more random variable's bit-representation are known.

Define a functional called \textit{information ambiguity} as ${\cal I}_{X_1, \ldots, X_N} = \lceil \log \mu_{X_1, \ldots, X_N} \rceil$ for a set of $N$ random variables $X_1, \ldots, X_N$. Next, we prove certain properties of functional ${\cal I}_{X_1, \ldots, X_N}$.

\begin{pavikl}
\label{lemma:Nexpansibility}
\textbf{(Expansibility)} If a component $(x_1, \ldots, x_N)$ with $p(x_1, \ldots, x_N) = 0$ is added to joint distribution $\mathcal P$, then information ambiguity ${\cal I}_{X_1,\ldots, X_N}$ does not change.
\end{pavikl}
\begin{IEEEproof}
The proof follows from the definition of $\mu_{X_1,\ldots, X_N}$.
\end{IEEEproof}

\begin{pavikl}
\label{lemma:Nmonotonicity}
\textbf{(Monotonicity)} If $A$ and $B$ are two discrete and finite sets with $A \subseteq B$, then ${\cal I}_A \le {\cal I}_B$.
\end{pavikl}
\begin{IEEEproof}
If $A \subseteq B$, then with $\mu_A = |A|$ and $\mu_B = |B|$
\begin{align*}
\mu_A & \le \mu_B \\
\log \mu_A & \le \log \mu_B \\
\lceil \log \mu_A \rceil & \le \lceil \log \mu_B \rceil
\end{align*}
Thus proving the lemma.
\end{IEEEproof}

\begin{pavikl}
\label{lemma:Nsymmetry}
\textbf{(Symmetry)} ${\cal I}_{X_1, \ldots, X_N} = {\cal I}_{\pi(X_1, \ldots, X_N)}$ for permutations $\pi(X_1, \ldots, X_N)$.
\end{pavikl}
\begin{IEEEproof}
The proof follows from the observation that any rearrangement of the elements of universal set $U$ does not change the cardinality of support-set $S_{X_1, \ldots, X_N}$.
\end{IEEEproof}

\begin{pavikl}
\label{lemma:Nsubadditivity}
\textbf{(Subadditivity)} If $X_i, 1 \le i \le N$, are ``interacting'', that is $S_{X_1, \ldots, X_N} \subseteq S_{X_1} \times \ldots \times S_{X_N}$, then  ${\cal I}_{X_1, \ldots, X_N} \le \sum_{i = 1}^N {\cal I}_{X_i}$.
\end{pavikl}
\begin{IEEEproof}
The proof follows from the straightforward extension of the proof of Lemma~\ref{lemma:2subadditivity} for multiple random variables.
\end{IEEEproof}

\begin{pavikl}
\label{lemma:Nadditivity}
\textbf{(Additivity)} If $X_i, 1 \le i \le N$, are ``non-interacting'', that is $S_{X_1, \ldots, X_N} = S_{X_1} \times \ldots \times S_{X_N}$, then ${\cal I}_{X_1, \ldots, X_N} \stackrel{\cdot}{=} \sum_{i = 1}^N {\cal I}_{X_i}$.
\end{pavikl}
\begin{IEEEproof}
Follows from extending the proof of Lemma~\ref{lemma:2additivity} to multiple random variables.
\end{IEEEproof}

The above lemmas establish that functional ${\cal I}_{X_1, \ldots, X_N} = \lceil \log \mu_{X_1, \ldots, X_N} \rceil$ is a valid information measure as it satisfies various axioms of expansibility, monotonicity, symmetry, subadditivity, and additivity of valid information measures \cite{106klir}.

\textit{Remark:} An astute reader may note that in spite of the apparent similarities between the information measures \textit{information ambiguity} proposed above and well-known Hartley measure, these two measures are fundamentally different. For a set of $N$ random variables these two information measures define their \textit{unconditional} versions identically in terms of functional $\log \mu_{X_1, \ldots, X_N}$. However, these two measures differ in their definitions of the corresponding \textit{conditional} versions. While conditional Hartley measure characterizes average nonspecificity, conditional information ambiguity characterizes maximum nonspecificity. For example, for a set of two random variables $X$ and $Y$, in conditional Hartley measure $H(\mu_X|\mu_Y) = \log \frac{\mu_{X_1, \ldots, X_N}}{\mu_Y}$, ratio $\frac{\mu_{X_1, \ldots, X_N}}{\mu_Y}$ represents the average number of elements of $\mu_X$ possible under the condition that an element from $\mu_Y$ has been chosen \cite[Chapter 2]{106klir}, while maximum conditional ambiguity $\widehat{\mu}_{X|Y}$ in conditional information ambiguity $\widehat{\cal I}_{X|Y}$ represents the maximum number of possible elements of $\mu_X$ under the condition that an element from $\mu_Y$ has been chosen.

\begin{pavikl}
\label{lemma:NambiguityBound}
Let $\Pi$ denote the set of all possible permutations of $\{1, \ldots, N\}$ and $\pi \in \Pi$, then
\begin{equation*}
{\cal I}_{X_1, \ldots, X_N} \le \min_{\pi \in \Pi} \sum_{i = 1}^N \widehat{\cal I}_{X_{\pi(i)}|X_{\pi(1)}, \ldots, X_{\pi(i-1)}}
\end{equation*}
\end{pavikl}
\begin{IEEEproof}
Combining the proofs of Lemma~\ref{lemma:2chainRule} and Corollary~\ref{corrl:2chainRule} generalized for $N$ random variables, proves the lemma. These generalizations themselves are easily obtainable from their two variables counterparts, as in the proof of Lemma~\ref{lemma:Nsubadditivity}.
\end{IEEEproof}

\begin{pavikl}
\label{lemma:ambiguitySet}
$S_{X_i|X_A}(x_A) = \bigcap_{X_j \in X_A} S_{X_i|X_j}(x_j)$
\end{pavikl}
\begin{IEEEproof}
We prove the lemma by individually proving both directions of inclusion.
\begin{itemize}
\item $S_{X_i|X_A}(x_A) \subset \bigcap_{X_j \in X_A} S_{X_i|X_j}(x_j)$: Consider $s \in S_{X_i|X_A}(x_A)$. We need to prove that $s \in S_{X_i|X_j}(x_j), X_j \in X_A$. By definition, $s$ is one of the values that the random variable $X_i$ can take when $X_j = x_j, \forall X_j \in X_A$. This implies that $s \in S_{X_i|X_j}(x_j), X_j \in X_A$.
\item $S_{X_i|X_A}(x_A) \supset \bigcap_{X_j \in X_A} S_{X_i|X_j}(x_j)$: Consider $s \in \bigcap_{X_j \in X_A} S_{X_i|X_j}(x_j)$. This implies that $s \in S_{X_i|X_j}(x_j), \forall X_j \in X_A$. Now, let us suppose that $s \not\in S_{X_i|X_A}(x_A)$. However, this leads to a contradiction as $S_{X_i|X_A}(x_A)$ is defined to be the set of all those values that $X_i$ can take, when $X_j = x_j, \forall X_j \in X_A$.
\end{itemize}
Combining these two proofs proves the lemma.
\end{IEEEproof}

\begin{pavikl}
\label{lemma:ambiguity}
$\mu_{X_i|X_A}(x_A) \le \min_{X_j \in X_A} \mu_{X_i|X_j}(x_j)$
\end{pavikl}
\begin{IEEEproof}
First consider the intersection of finite number of finite sets $A_i, i \in I$, where $I$ is some index set.
\begin{align*}
\mu_{X_i|X_A}(x_A) & \stackrel{(a)}{=} |S_{X_i|X_A}(x_A)| \\
                   & \stackrel{(b)}{=} |\bigcap_{X_j \in X_A} S_{X_i|X_j}(x_j)| \\
                   & \stackrel{(c)}{\le} \min_{X_j \in X_A} |S_{X_i|X_j}(x_j)| \\
                   & = \min_{X_j \in X_A} \mu_{X_i|X_j}(x_j),
\end{align*}
where (a) follow from the definition $\mu_{X_i|X_A}(x_A)$, (b) follows from the Lemma~\ref{lemma:ambiguitySet}, and (c) follows from $|\bigcap_{i \in I} A_i| \le \min_{i \in I} |A_i|$. This proves the lemma.
\end{IEEEproof}

\begin{pavikl}
\label{lemma:maxAmbiguity}
$\widehat{\mu}_{X_i|X_A} \le \min_{X_j \in X_A} \widehat{\mu}_{X_i|X_j}$
\end{pavikl}
\begin{IEEEproof}
From the definition of $\widehat{\mu}_{X_i|X_A}$, let $x_A^*$ be an instance of $X_A$ that maximizes $\mu_{X_i|X_A}(x_A)$. Similarly, by the definition of $\widehat{\mu}_{X_i|X_j}, X_j \in X_A$, let $x_j^{'}$ be an instance of $X_j$ that maximizes $\mu_{X_i|X_j}(x_j), X_j \in X_A$. Therefore,
\begin{align*}
\widehat{\mu}_{X_i|X_A} & = \mu_{X_i|X_A}(x_A^*) \\
                       & \stackrel{(a)}{\le} \min_{X_j \in X_A} \mu_{X_i|X_j}(x_j^*) \\
                       & \le \min_{X_j \in X_A} \mu_{X_i|X_j}(x_j^{'}) \\
                        & = \min_{X_j \in X_A} \widehat{\mu}_{X_i|X_j},
\end{align*}
where (a) follows from Lemma~\ref{lemma:ambiguity}, thus completing the proof.
\end{IEEEproof}

\section{Notation}
This section provides the notation used frequently in rest of the paper.
\begin{enumerate}
\item [$\mathcal{N}$:] the set of $N$ informants.
\item [$\mathcal X$:] finite, discrete alphabet set of size $|{\mathcal X}|$.
\item [${\mathcal P}$:] $N$-dimensional discrete probability distribution, ${\cal P}$ $= p(x_1,$ $ \ldots, x_N), x_i \in {\mathcal X}$.
\item [$X_i$:] random variable observed by the informant $i$. $X_i \in {\cal X}$.
\item [$S_{X_i}$:] the \textit{ambiguity set} of the $i^\textrm{th}$ informant's data, with corresponding \textit{ambiguity} $\mu_{X_i} = |S_{X_i}|$.
\item [$S_{X_{i}|I}$:] the \textit{conditional ambiguity set} of the sink in the $i^{\textrm{th}}$ informant's data when the sink has information $I$, which can be the set of values of one or more bit-locations in the representation of one or more informants' data. However, the exact nature of $I$ will be obvious from the context.
\item [$\mu_{X_{i}|I}$:] the \textit{conditional ambiguity}, $|S_{X_{i}|I}|$.
\item [$\widehat{\mu}_{X_{i}|I}$:] the \textit{maximum conditional ambiguity}, computed over all instances of $I$.
\item [$S_{X_1, \ldots, X_N}$:] the \textit{ambiguity set} at the sink of all informants' data, with $\mu_{X_1, \ldots, X_N} = |S_{X_1, \ldots, X_N}|$ as the corresponding \textit{ambiguity}.
\item [$S_{X_1, \ldots, X_N|I}$:] the \textit{conditional ambiguity set} at the sink of all informants' data, with $\mu_{X_1, \ldots, X_N|I} = |S_{X_1, \ldots, X_N|I}|$ as the corresponding \textit{conditional ambiguity}.
\item [$S^k_{X_i}$:] the $k^{\mathrm{th}}$-extension of ambiguity set $S_{X_i}, i \in {\mathcal N}$, with corresponding \textit{ambiguity} $\mu^k_{X_i} = |S^k_{X_i}|$.
\item [$S^k_{X_i|I}$:] the \textit{conditional} $k^{\textrm{th}}$-extension of ambiguity set $S_{X_i}$ when the sink has information $I$, with corresponding \textit{conditional ambiguity} $\mu^k_{X_i|I} = |S^k_{X_i|I}|$.
\item [$S^k_{X_1, \ldots, X_N}$:] the $k^{\mathrm{th}}$-extension of ambiguity set $S_{X_1, \ldots, X_N}$, with $\mu^k_{X_1, \ldots, X_N} = |S^k_{X_1, \ldots, X_N}|$ as the corresponding \textit{ambiguity}.
\item [$S^k_{X_1, \ldots, X_N|I}$:] the \textit{conditional} $k^{\textrm{th}}$-extension of ambiguity set $S_{X_1, \ldots, X_N}$ when the sink has information $I$, with corresponding \textit{conditional ambiguity} $\mu^k_{X_1, \ldots, X_N|I} = |S^k_{X_1, \ldots, X_N|I}|$.
\item [$C_B$:] the worst-case \textit{bit-compressibility} of distributed compression problem with single instance of source data-vector.
\item [$C_B^k$:] the worst-case \textit{bit-compressibility} of distributed compression problem with $k, k>1$, instances of source data-vectors.
\end{enumerate}

\section{Problem Setting}
\label{sec:probSetting}
Consider a distributed information-gathering scenario, where a sink collects the data from $N$ informants sampling correlated data. Divide the sequence of events in this data-gathering problem in terms of \textit{data-generation epoch} and \textit{data-gathering epoch}. In the data-generation epoch, a sample $\overline{x} = (x_1, \ldots, x_N), \overline{x} \in S_{X_1, \ldots, X_N}$, is drawn from the discrete and finite support-set $S_{X_1, \ldots, X_N}$ over $N$ binary strings, as in \cite{104chouPetrovic, 105adler}. The strings of $\overline{x}$ are revealed to the informants, with the string $x_i$ being given to the $i^\textrm{th}$ informant, $i \in \mathcal{N}$. Then in the data-gathering epoch, the sink wants to \textit{losslessly} (Error probability $P_e = 0$) learn $\overline{x}$ revealed to the informants. Each data-generation epoch is followed by a data-gathering epoch and vice-versa.

\begin{figure}[!t]
\centering
\includegraphics[width=4.0in]{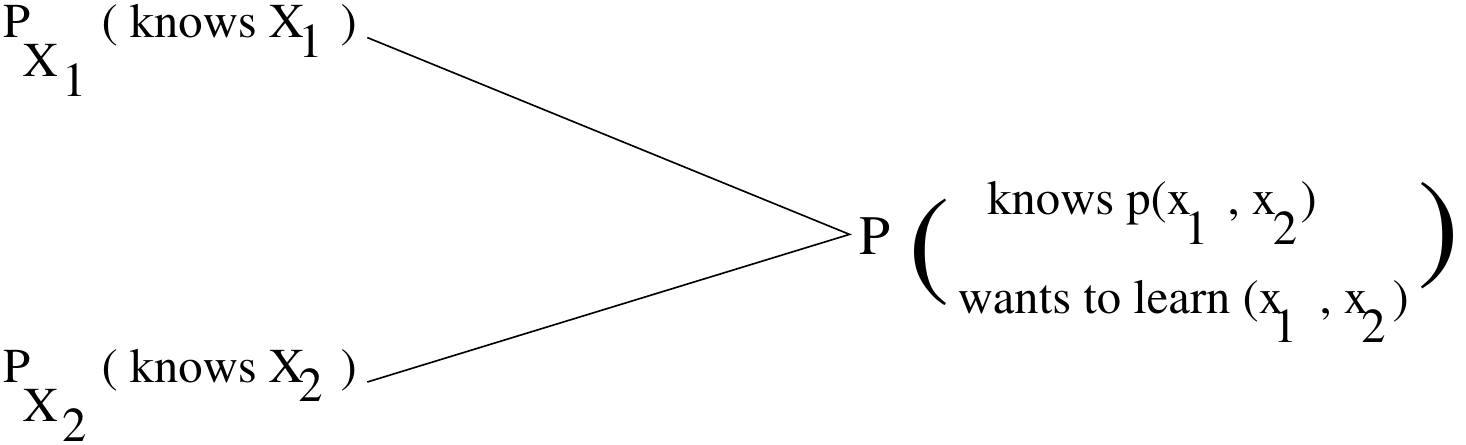}
\caption{``Two informants - Single sink'' communication problem.}
\label{fig:twoInformants}
\end{figure}

\textbf{Problem Statement:} A sample $\overline{x} = (x_1, \ldots, x_N)$ is drawn \textit{i.i.d.} from the distribution $\mathcal P$ over $N$ binary strings. The strings of $\overline{x}$ are revealed to the informants, with the string $x_i$ being given to the $i^\textrm{th}$ informant. The sink wants to learn each informant's string \textit{losslessly} ($P_e = 0$). An informant may not learn about other informants' or the sink's data. Our primary objective is to minimize the total number of informant bits required, in the worst-case, to accomplish this, but we are also concerned with minimizing both, the number of rounds and the number of sink bits. This is illustrated in Figure~\ref{fig:twoInformants} for the scenarios with two informants and one sink.

\textbf{The Problem Setting:} We consider an \textit{asymmetric communication} scenario\footnote{Our formalism can also be applied to the communication scenarios, where even the sink does not know $\mathcal P$, but can estimate it as it collects the data from the informants, drawn from $\mathcal P$. For example, in \cite{104chouPetrovic}, a linear predictive model is used to estimate the correlation structure. It should be noted that we assume nothing about this distribution, except that it is a discrete distribution with finite alphabet.} \cite{097kushilevitzNisan}. Communication takes place over $N$ binary, error-free channels, where each channel connects an informant with the sink. An informant and the sink can interactively communicate over the channel between them by exchanging messages (finite sequences of bits determined by an agreed upon, deterministic protocol.) The informants cannot communicate directly with each other. We assume that in the data-gathering epoch, communication between the sink and the informants proceeds in rounds, as in \cite{079yao}. In each round, depending on the information held by the communicators, one or other communicator may send the first message. However, as argued in \cite{090orlitsky}, if we allow the empty messages and eliminate the last message if it is sent by the sink, then any sequence of messages can be converted into another sequence where the same communicator transmits the first message, with no increase in the worst-case communication complexity. Therefore, we assume that in each communication round, first the sink communicates to the informants and then, the informants respond with their messages. Each bit communicated over any channel is counted as either a sink bit or an informant bit.

We assume the informants to be memoryless in the sense that they do not remember their messages sent in the previous rounds. However, we assume that the $i^\textrm{th}$ informant knows its support-set $S_{X_i}$, so that it represents the binary string $x_i$ given to it in ${\cal I}_{X_i} = \lceil \log \mu_{X_i} \rceil$ bits as $b_1^i \ldots b_{\lceil \log \mu_{X_i} \rceil}^i$.

The sink knows distribution $\mathcal P$ and the corresponding support set $S_{X_1, \ldots, X_N}$. So, every $\overline{x}, \overline{x} \in S_{X_1, \ldots, X_N}$, can be uniquely described at the sink using ${\cal I}_{X_1, \ldots, X_N} = \lceil \log \mu_{X_1, \ldots, X_N} \rceil$ bits. This implies that, in the worst-case, ${\cal I}_{X_1, \ldots, X_N}$ informant bits are \textit{necessary} for the sink to learn $\overline{x}$ unambiguously. However, these many informant bits may not be achievable, in general, for any communication protocol as the sink needs to query the informants based on some function of independent encoding of their data-strings that the informants can construct rather than some arbitrary encoding of $\overline{x}$ that the sink can construct. However, as long as the sink can be assumed to know joint distribution $\mathcal P$, there is at least one coding scheme that both, the sources and sink can construct without any explicit communication between them and still achieve optimal distributed compression performance. Next, we propose one such encoding scheme that the sink can construct to query the informants and informants can use to respond to the sink's queries. This scheme allows us to not only compute minimum achievable number of informant bits required for data-gathering at the sink but also provides an efficient way to achieve those.

\textbf{New Problem Encoding Scheme:} As each informant $i, i \in \mathcal{N}$, knows its support-set $S_{X_i}$, it can describe each $x_i, x_i \in S_{X_i}$, as set $B_{x^i}$ of ${\cal I}_{X_i}$ bits\footnote{This encoding scheme can be agreed upon \textit{a priori} between the sink and and each informant.}. Therefore, every $\overline{x}$ can also be uniquely described at the sink as set $B_{\overline{x}}$ of $\sum_{i \in \mathcal{N}} {\cal I}_{X_i}$ bits, constructed by concatenating ${\cal I}_{X_i}$ bits long representation of each $x_i, i \in \mathcal{N}$. This implies that $\sum_{i \in \mathcal{N}} {\cal I}_{X_i}$ informant bits are always \textit{sufficient} for the sink to learn $\overline{x}$ unambiguously. The following example illustrates this encoding scheme.

\begin{figure*}[!t]
\centering
\includegraphics[width=5.5in]{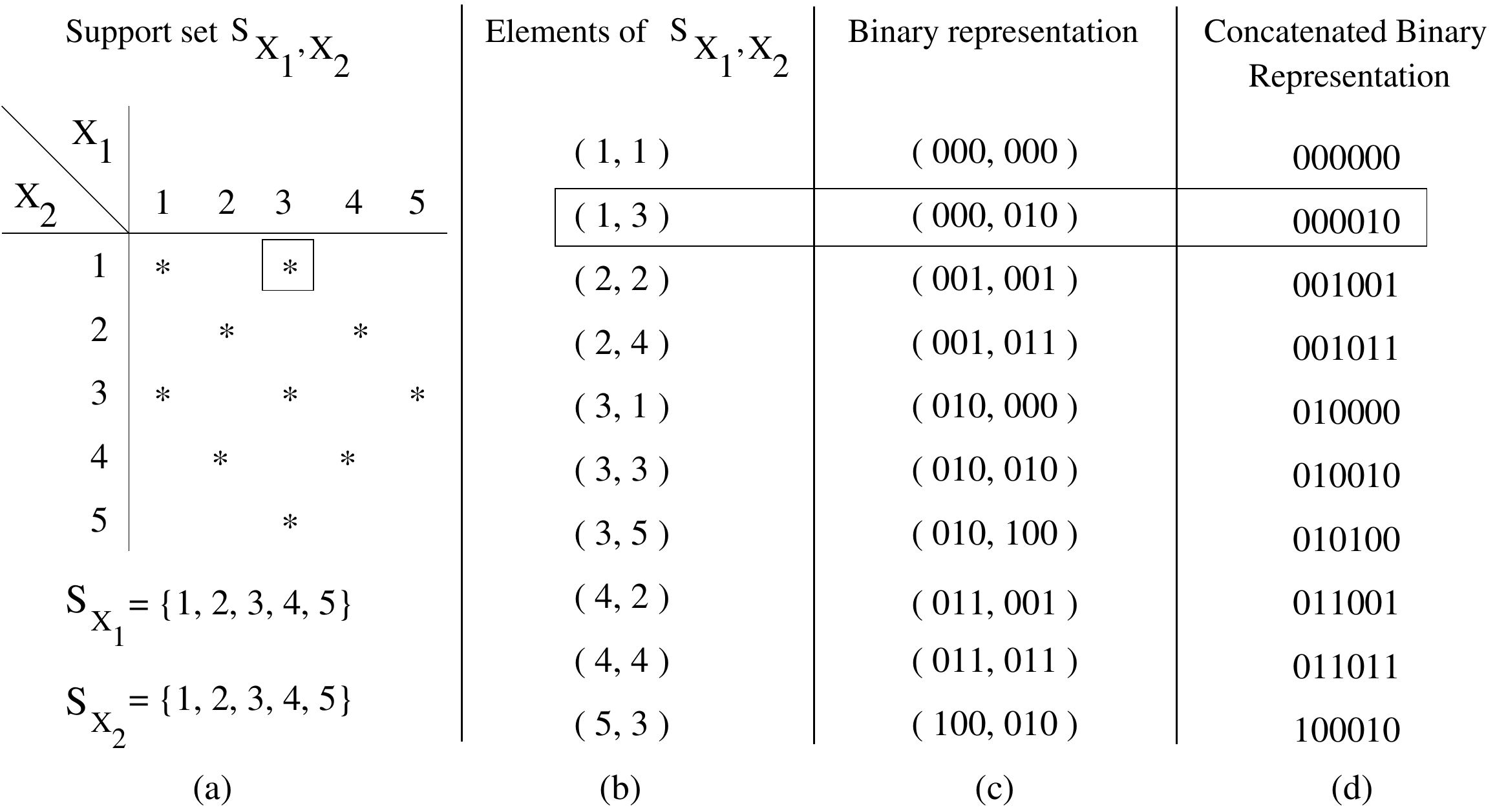}
\caption{Example of problem setting: (a) Support-sets: $S_{X_1, X_2}, S_{X_1}, S_{X_2}$ with $\mu_{X_1, X_2} = 10, \mu_{X_1} = \mu_{X_2} = 5$ (b) the members of $S_{X_1, X_2}$ (c) binary representation of members of $S_{X_1, X_2}$ (d) the concatenated binary representation. If the string `$000010$' is drawn, then `$000$' is given to informant 1 and `$010$' is given to informant 2.}
\label{fig:probExa}
\end{figure*}

\textbf{Example 1:} Consider an example support-set shown in Figure~\ref{fig:probExa}. Let informants 1 and 2 observe the random variables $X_1$ and $X_2$, respectively. For the given support-set, at least ${\cal I}_{X_1, X_2} = 4$ bits are required to describe any element of $S_{X_1, X_2}$ and it requires no less than $3$ bits to independently describe a value that $X_1$ or $X_2$ take.

For a given support-set, the sink can construct a figure similar to Figure~\ref{fig:probExa}. One of the strings from the fourth column is drawn, with first ${\cal I}_{X_1}$ bits given to informant 1, next ${\cal I}_{X_2}$ bits given to informant 2, and so on. Then the data-gathering problem is that the sink wants to learn of this string, whose different parts are held by different informants, with the informants sending minimum total number of bits to the sink. \endproof

Given the above encoding scheme, the worst-case \textit{bit-compressibility} $C_B$ of distributed compression problem is defined as:
\begin{equation}
\label{eqn:bitCompressibility}
C_B = \max_{\overline{x} \in S_{X_1, \ldots, X_N}} \min_{b_{\overline{x}} \subseteq B_{\overline{x}}} |b_{\overline{x}}| \mbox{ such that } S_{X_1, \ldots, X_N|b_{\overline{x}}} = \{\overline{x}\},
\end{equation}
where $S_{X_1, \ldots, X_N|b}$ denotes the conditional ambiguity-set of $(X_1, \ldots, X_N)$ when the bits corresponding to subset $b_{\overline{x}}, b_{\overline{x}} \subseteq B_{\overline{x}}$, are known at the sink\footnote{Note that $b_{\overline{x}}$ can be written as concatenation of $b_{x^i} \subseteq B_{x^i}, i \in \mathcal{N}$.}. In other words, there is at least one $\overline{x} \in S_{X_1, \ldots, X_N}$ such that no fewer than $C_B$ informant bits are sufficient to describe it at the sink unambiguously. Note that ${\cal I}_{X_1, \ldots, X_N} \le C_B \le \sum_{i \in N} {\cal I}_{X_i}$.

\textbf{Definition (Worst-case Bit-Compressibility):}
The distributed compression problem is called worst-case bit-compressible if $C_B < \sum_{i \in N} {\cal I}_{X_i}$, otherwise bit-incompressible.

\textit{Note on the terminology:} We call a bit \textit{undefined} if the sink does not know its value, otherwise it is called \textit{defined}. For example, until the sink learns of actual $\overline{x}$ revealed to the informants, one or more bits in the $\sum_{i = 1}^N {\cal I}_{X_i}$ bits long representation of $\overline{x}$, remain \textit{undefined}. Similarly, informant $i, i \in \mathcal{N}$, is called \textit{undefined} if sink does not know corresponding $x_i, x_i \in \overline{x}$ exactly, otherwise the informant is called \textit{defined}.

\section{Bit-compressibility for Distributed Compression}
\label{sec:bCompressibility4cmprsn}
In this section, we address the problem of worst-case distributed compression in two different communication scenarios. In the first scenario that we call bit serial communication scenario, in each communication round in a data-gathering epoch, only one informant can send one bit of information to the sink. This allows us to compute the minimum number of informant bits (total and individual) required to enable the sink to learn the particular $\overline{x}, \overline{x} \in S_{X_1, \ldots, X_N}$, revealed to the informants in the data-generation epoch, when any number of rounds and sink bits can be used. In other words, this communication scenario allows us to compute the largest worst-case achievable rate-region for this problem, as we show later. In the second scenario that we call round parallel communication scenario, one or more informants can send one or more bits in parallel to the sink. This as we argue and show later, allows us to exploit various trade-offs among the number of informant bits, the number of sink bits, and the number of rounds.

\textbf{Definition (Achievable Rate-Region):}
The achievable rate-region $\mathcal R$ for the worst-case distributed source coding problem with $N$ informants is defined as the set of all $N$-tuples $(b_{x^1}, \ldots, b_{x^N})$ of informant rates (in bits) such that when $i^\textrm{th}, i \in \mathcal{N}$, informant sends the subset $b_{x^i}, b_{x^i} \subseteq B_{x^i}$, of bits in the particular rate-tuple, then the sink is able to retrieve $\overline{x}$ unambiguously, that is:
\begin{equation}
\label{eqn:rateRegionDef}
\mathcal{R} = \{(b_{x^1}, \ldots, b_{x^N})| S_{X_1, \ldots, X_N|(b_{x^1}, \ldots b_{x^N})} = \{\overline{x}\} \mbox{ and } b_{x^i} \subseteq B_{x^i}, i \in \mathcal{N}\}
\end{equation}

\subsection{Bit Serial Communication}
\label{subsec:bSerCom}
We discuss the optimal solution of the distributed compression problem introduced in the previous section. We first provide an interactive communication protocol, called ``\textbf{Bit-Serial}'' protocol, and then prove that it optimally solves the problem. Further, we show that ``\textbf{Bit-Serial}'' protocol also allows us to compute the maximum achievable rate-region of the distributed compression problem we are concerned with. Next, we describe ``\textbf{Bit-Serial}'' protocol in detail.

\textit{\textbf{Bit-Serial} Protocol:} Consider an interactive communication protocol where in each round only one bit is sent by the informant chosen to communicate with the sink. The chosen bit has the property that it divides the current conditional ambiguity set at the sink maximally close to half among all candidate bits. This offers the opportunity to optimally minimize the number of informant bits, as it maximally conditions the ambiguity sets of the informants at the sink.

Consider the $l^\textrm{th}$ communication round. At the beginning of the $l^\textrm{th}$ round, let $U$ and $D$ denote, respectively, the sets of \textit{undefined} and \textit{defined} bit-locations among $\sum_{i = 1}^{N} {\cal I}_{X_i}$ bits long representation of $\overline{x}$ at the sink, $|U| + |D| = \sum_{i = 1}^{N} {\cal I}_{X_i}$. The ambiguity at the sink in all informants' data is $\mu^l_{X_1,\ldots, X_N} = \mu_{{X_1,\ldots, X_N}|D}$. Let $N_i^0, N_i^1$ respectively denote the number of $0s$ and $1s$ at the bit location $i \in U$, over all $\mu^l_{X_1,\ldots, X_N}$ strings. Then the chosen bit is the one that solves $\argmin_{i \in U} |N_i^0 - N_i^1|$. The sink, after receiving the value of the chosen bit, recomputes the set of undefined bits $U$. This is carried out iteratively till all bits in $\sum_{i = 1}^N {\cal I}_{X_i}$ bits long representation of $\overline{X}$ are not \textit{defined}. This is formally summarized in ``\textbf{Bit-Serial}'' protocol given below.

\hspace{-1.0em}\hrulefill

\hspace{-0.5em}{\textbf{Protocol:} Bit-Serial}

\vspace{-0.2cm}\hspace{-1.0em}\hrulefill
\begin{codebox}
\li $l = 0$
\li Let $S_{X_1,\ldots, X_N}^l = S_{X_1,\ldots, X_N}$
\li Let $V = \{1, \ldots, \sum_{i = 1}^{N} {\cal I}_{X_i}\}$: index set of all bit-locations in $B_{\overline{x}}$
\li Let $U$ be the index set of undefined bits in $V$, $U \subseteq V$
\li \While ($\mu_{X_1,\ldots, X_N}^l > 1$)
\li \hspace{-0.45in} $J^{l+1} = \argmin_{i \in U} |N_i^0 - N_i^1|$
\li \hspace{-0.45in}  If $|J^{l+1}| > 1$, then choose uniformly at random the bit-location  $j^{l+1}, j^{l+1} \in J^{l+1}$
\li \hspace{-0.45in}  The sink asks the informant corresponding to bit-location $j^{l+1}$ to send bit-value $b(j^{l+1})$ \label{bSerCom:bitValue}
\li \hspace{-0.45in}  Set $S_{X_1,\ldots, X_N}^{l+1} = S_{{X_1,\ldots, X_N}|b(j^{l+1})}^l$
\li \hspace{-0.45in}  Compute $U \subset V$, the set of undefined bits
\li \hspace{-0.45in}  $l = l + 1$
    \End
\end{codebox}
\vspace{-0.2cm}\hrulefill

The sink can perform the worst-case performance analysis of \textbf{Bit-Serial} protocol by selecting on the Line~\ref{bSerCom:bitValue}, $b^*(j^{l+1})$ that solves:
\begin{equation*}
b^*(j^{l+1}) = \argmax\limits_{s = \{0, 1\}} \mu^l_{{X_1,\ldots, X_N}|b(j^{l+1})=s}
\end{equation*}

\begin{figure*}[!t]
\centering
\includegraphics[width=5.25in]{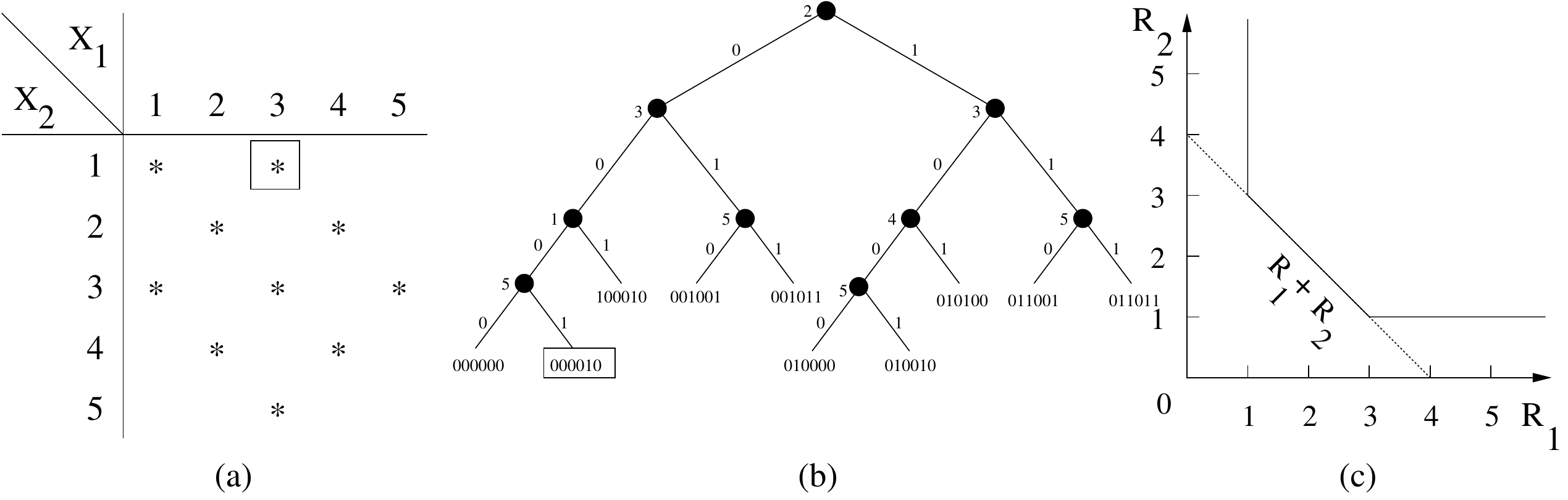}
\caption{Example 1: (a) Support set $S_{X_1, X_2}$, with $\mu_{X_1, X_2} = 10, {\cal I}_{X_1, X_2} = 4$ (b) one of the minimum-height binary trees generated by \textbf{Bit-Serial} protocol corresponding to $S_{X_1, X_2}$. The number appearing on the left of every node corresponds to the bit-location in the concatenated binary representation, as in Figure~\ref{fig:probExa}(d), of the elements of $S_{X_1, X_2}$ (c) worst-case achievable rate-region. The string `$000010$', drawn as in Figure~\ref{fig:probExa}, is highlighted.}
\label{fig:exa1}
\end{figure*}

The binary representations of elements of $S_{X_1, \ldots, X_N}$ in terms of $B_{\overline{x}}$, as in Figure~\ref{fig:probExa}(d), can be arranged as the leaves of a binary tree. For each of $\sum_{i = 1}^{N} {\cal I}_{X_i}$ bit-locations in the $B_{\overline{x}}$ representation of $\overline{x}$, there is a binary tree rooted at that location with all other locations forming the internal nodes of the tree. At any node in the tree, the bit-value `$0$' leads to the left subtree and `$1$' leads to the right subtree. Such a binary tree with $\mu_{X_1, \ldots, X_N}$ leaves will have a minimum-height of ${\cal I}_{X_1, \ldots, X_N}$, implying that at least ${\cal I}_{X_1, \ldots, X_N}$ bits are required to describe any leaf, in the worst-case. Figure~\ref{fig:exa1} provides the canonical representation of one of the possible binary trees for the distributed compression problem in Figure~\ref{fig:probExa}.

We show that the problem of minimizing the total number of bits $C_B$ that the informants must send to the sink to help it learn any $\overline{x} \in S_{X_1, \ldots, X_N}$ is equivalent to the problem of constructing minimum-height binary tree for concatenated bit-representations of the elements of $S_{X_1, \ldots, X_N}$. We prove that \textbf{Bit-Serial} protocol constructs such trees for a given support set and so optimally solves the worst-case asymmetric distributed compression problem.

\begin{pavikl}
\label{lemma:bSerComAllTrees}
\textbf{Bit-Serial} protocol computes all minimum-height binary trees corresponding to the given support-set.
\end{pavikl}
\begin{IEEEproof}
In the canonical representation, as in Figure~\ref{fig:exa1}, of a minimum-height binary tree corresponding to the given support-set, every node corresponds to the bit-location that divides the resultant conditional ambiguity set as close to half as possible. However, \textbf{Bit-Serial} protocol precisely chooses the same bit-location in the round corresponding to the level of node concerned, thus proving the lemma.
\end{IEEEproof}

Denote the set of all minimum-height binary trees as $\mathcal T$. Let $b_i^j$ denote the number of bits that the $i^{\textrm{th}}$ informant, $i \in \mathcal{N}$, sends in the worst-case under the $j^{\textrm{th}}$ minimum-height binary tree, $j \in \mathcal{T}$. Then, we have the following lemma.

\begin{pavikl}
\label{lemma:bSerComMinBits}
\textbf{Bit-Serial} protocol computes $b_i$, the minimum number of bits that the $i^{\textrm{th}}$ informant needs to send to the sink to get defined.
\end{pavikl}
\begin{IEEEproof}
\textbf{Bit-Serial} protocol exploits the bit serial communication scenario where the informant chosen to communicate in a round can send only one bit of information to maximally condition the resultant ambiguity set at the sink. Also, to reduce the number of bits that an informant sends, \textbf{Bit-Serial} protocol can postpone retrieving the bits from the informant concerned until it can be postponed no more, thus maximally conditioning the ambiguity set at the sink of the informant concerned. These two arguments  together prove the lemma.
\end{IEEEproof}

Combining previous two lemmas, allows us to define $b_i$ as:
\begin{equation}
\label{eqn:minInformantBits}
b_i = \min_{j \in \mathcal{T}} b_i^j,
\end{equation}

\begin{pavikl}
\label{lemma:rateRegion}
For a given support-set, each corner point of the worst-case achievable rate-region corresponds to at least one minimum-height binary tree, with height $C_B$.
\end{pavikl}
\begin{IEEEproof}
For the sake of contradiction, assume that there is a corner point of the worst-case achievable rate-region to which no minimum-height binary tree corresponds to. This means that this corner point is outside the worst-case rate-region defined by the set of all the corner points visited by the set of minimum-height binary trees, $\mathcal{T}$. This further implies that at this corner point at least one informant, say the $i^{\textrm{th}}$, sends fewer bits than $b_i$ with $b_i$ as defined above. However, this contradicts the definition of $b_i$ that it is the minimum number of bits an informant needs to send to the sink before it is defined. Thus, there cannot be any corner point outside the rate-region defined by the set of corner points defined by the minimum-height binary trees in $\mathcal{T}$, hence proving the lemma.
\end{IEEEproof}

\begin{pavikl}
\label{lemma:bSerComOptimal}
Protocol \textbf{Bit-Serial} is worst-case optimal.
\end{pavikl}
\begin{IEEEproof}
Combining the statements of Lemmas~\ref{lemma:bSerComAllTrees} and \ref{lemma:rateRegion}, we can state that \textbf{Bit-Serial} protocol computes at least one minimum-height binary tree corresponding to each corner point of the worst-case achievable rate-region. Therefore, \textbf{Bit-Serial} protocol computes each corner point of the achievable rate-region. Thus, \textbf{Bit-Serial} protocol computes the worst-case achievable rate-region, hence it is worst-case optimal.
\end{IEEEproof}

For two informants, the worst-case achievable rate-region in asymmetric distributed compression problem is given by the following corollary to Lemma~\ref{lemma:bSerComOptimal}.

\begin{pavikc}
\label{cor:2rateRegion}
For $N = 2$, if $b_i$ denotes the minimum number of bits that an informant $i, 1 \le i \le 2$, sends over all solutions of \textbf{Bit-Serial} protocol and $C_B$ denotes the minimum total number of bits sent by all informants, then the achievable rate region is given by:
\begin{align*}
R_1 & \ge b_1 \\
R_2 & \ge b_2 \\
R_1 + R_2 & \ge C_B
\end{align*}
\end{pavikc}
\begin{IEEEproof}
Follows from the worst-case optimality of \textbf{Bit-Serial} protocol proven above.
\end{IEEEproof}

For $N$ informants, the worst-case achievable rate-region in asymmetric distributed compression problem is given by the following corollary to Lemma~\ref{lemma:bSerComOptimal}.

\begin{pavikc}
\label{cor:NrateRegion}
The set of achievable rate-vectors for the worst-case DSC problem for the oneshot compression is given by: $R(S) \ge M_{R(S)}$ for $S, S \subseteq \mathcal{N}$, where $M_{R(S)}$ is the minimum number of bits that the subset of informants $S$ send over all possible solutions of \eqref{eqn:bitCompressibility} and $R(S) = \sum_{i \in S} R_i$.
\end{pavikc}
\begin{IEEEproof}
The proof follows from establishing the worst-case optimality of \textbf{Bit-Serial} protocol in computing $M_{R(S)}$. This can be achieved by the straightforward generalization of the argument above to prove the optimality of \textbf{Bit-Serial} protocol to arbitrary subsets of $S, S \subseteq \mathcal{N}$.
\end{IEEEproof}

\begin{figure*}[!t]
\centering
\includegraphics[width=5.25in]{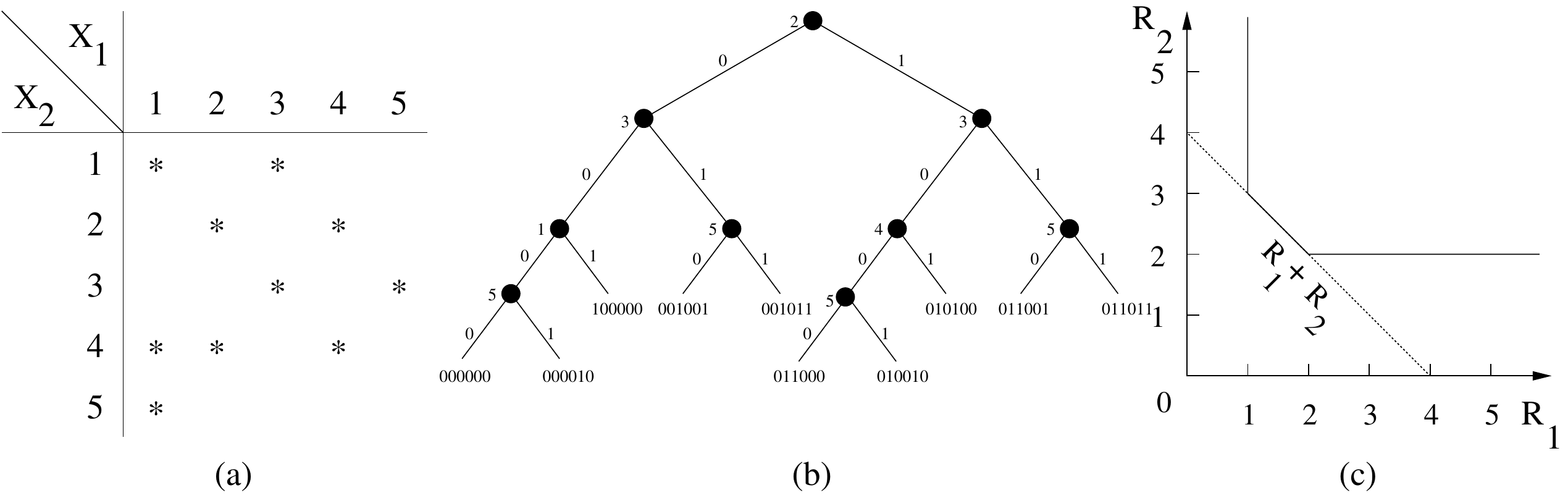}
\caption{Example 2: (a) Support set $S_{X_1, X_2}$, with $\mu_{X_1, X_2} = 10, {\cal I}_{X_1, X_2} = 4$ (b) one of the minimum-height binary trees generated by \textbf{Bit-Serial} corresponding to $S_{X_1, X_2}$ (c) worst-case achievable rate-region}
\label{fig:exa2}
\end{figure*}

\begin{figure*}[!t]
\centering
\includegraphics[width=5.25in]{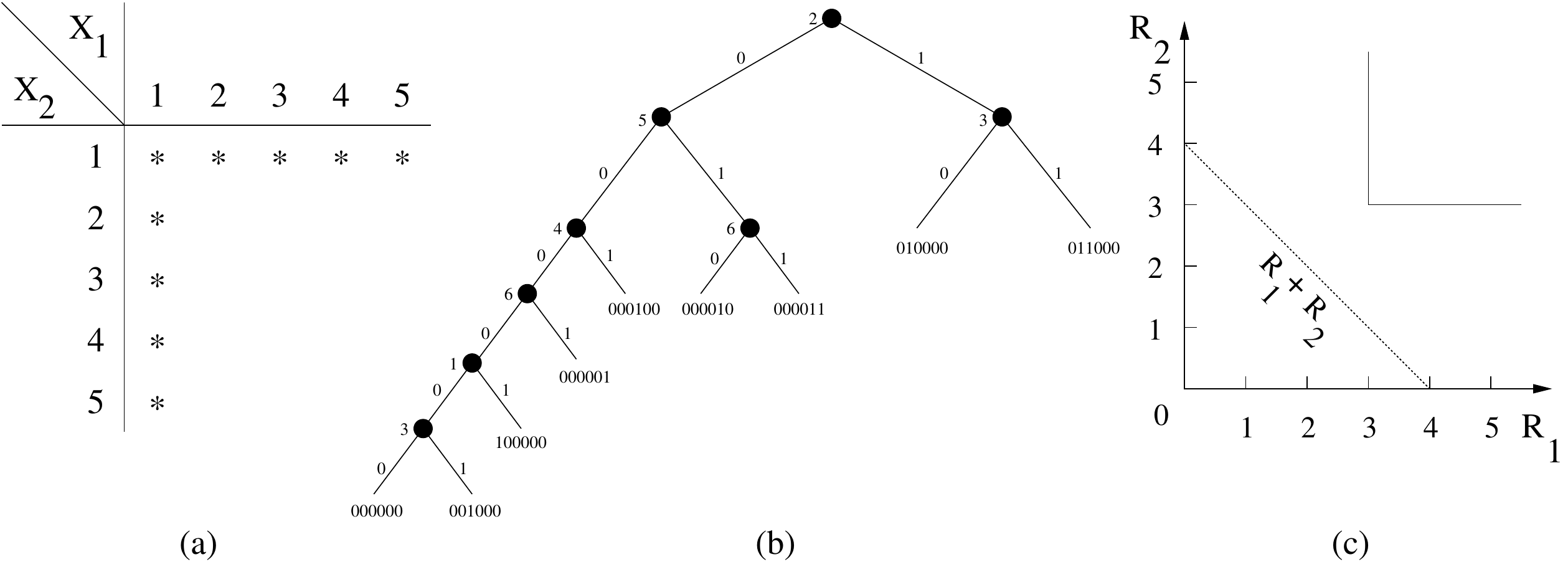}
\caption{Example 3: (a) Support set $S_{X_1, X_2}$, with $\mu_{X_1, X_2} = 9, {\cal I}_{X_1, X_2} = 4$ (b) one of the minimum-height binary trees generated by \textbf{Bit-Serial} corresponding to $S_{X_1, X_2}$ (c) worst-case achievable rate-region}
\label{fig:exa3}
\end{figure*}

In Figures~\ref{fig:exa1}-\ref{fig:exa3}, for three different support sets, we give one of the many possible corresponding minimum-height trees computed by \textbf{Bit-Serial} protocol and the corresponding worst-case achievable rate regions.

\textit{Upper bound on $C_B$}: In \textbf{Bit-Serial} protocol, as only one information bit is sent per communication round, the total number of rounds required is equal to $C_B$. Assume that in the $i^\textrm{th}, 1 \le i\le C_B$, communication round, the size of the ambiguity set is reduced by $2^{(1-\epsilon_i)}, 1 - {\cal I}_{X_1,\ldots, X_N} < \epsilon_i < 1$. So, after $C_B$ rounds, we have $\frac{\mu_{X_1,\ldots, X_N}}{2^{\sum_{s=1}^{C_B} (1-\epsilon_s)}} = 1$.

Define $\epsilon = \max\{\epsilon_1, \ldots, \epsilon_k\}$. Assume that the size of the ambiguity set in every round is reduced by $2^{(1-\epsilon)}$. Assume that the data-gathering finishes now in $k$ rounds. It is obvious that $C_B \le k$. Now, the size of the ambiguity set after $k$ rounds satisfies $\frac{\mu_{X_1,\ldots, X_N}}{(2^{(1-\epsilon)})^k} = 1$. This implies that $k = \big\lceil \frac{\log \mu_{X_1,\ldots, X_N}}{(1-\epsilon)}\big\rceil$.

\textit{Upper bound on number of sink bits}: As there are $N$ informants, under \textbf{Bit-Serial} protocol, in the $i^\textrm{th}$ communication round, the sink addresses the chosen informant in $\lceil \log N \rceil$ bits and then in $\lceil \log \log \mu_{X_i} \rceil$ bits addresses the chosen bit corresponding to this informant. So, in the $i^\textrm{th}$ communication round, the sink sends a total of $\lceil \log N \rceil + \lceil \log \log \mu_{X_i} \rceil$ bits, implying that to gather $B$ information bits the sink sends a total of $C_B\lceil \log N \rceil + \sum_{i = 1}^{C_B} \lceil \log \log \mu_{X_i} \rceil$ bits.

\subsection{Round Parallel Communication}
\label{subsec:rParCom}
In this subsection, we investigate the worst-case performance of the interactive distributed compression problem in an asymmetric communication scenarios where in each communication round one or more informants may send more than one bit to the sink \textit{in parallel}. More precisely, we use parallel communication to mean \textbf{round parallel communication} that we define as a communication scenario where in each communication round, two or more bits can be sent by one or more informants to the sink. Therefore, in the round parallel communication two or more informants may or may not communicate in parallel in the classical sense, that is, their communications may or may not overlap in time.

The round parallel communication allows us to exploit the trade-off between the number of rounds and the number of informant bits. Therefore, on one extreme is \textbf{Bit-Serial} protocol with minimum number of informant bits and unconstrained number of rounds and on other extreme is a scheme where as many as $\sum_{i=1}^N {\cal I}_{X_i}$ informant bits are sent (as each informant $i$ encodes its data-value in ${\cal I}_{X_i}$ bits) in a single round.

We provide a round parallel communication protocol that, as we prove, among all round parallel communication protocols minimizes the total number of informant bits in the worst-case as well as the number of sink bits and the number of rounds.

\textit{\textbf{Round-Parallel} Protocol:} Consider the set of concatenated bit-strings of length $\sum_{i = 1}^N {\cal I}_{X_i}$, where each bit-string corresponds to an element of $S_{X_1,\ldots, X_N}$, as in Figure~\ref{fig:probExa}(d). Consider the $l^\textrm{th}$ communication round. At the beginning of the $l^\textrm{th}$ round, let $U$ and $D$ denote, respectively, the sets of \textit{undefined} and \textit{defined} bit-locations among $\sum_{i = 1}^{N} {\cal I}_{X_i}$ bits, $|U| + |D| = \sum_{i = 1}^{N} {\cal I}_{X_i}$. The ambiguity at the sink in all informants' data is $\mu^l_{X_1,\ldots, X_N} = \mu_{{X_1,\ldots, X_N}|D}$. Let $N_i^0, N_i^1$ respectively denote the number of $0s$ and $1s$ at the bit location $i \in U$, over all $\mu^l_{X_1,\ldots, X_N}$ strings. The sink computes $J^l$, the set of indices of those $\lceil \log \mu^l_{X_1,\ldots, X_N} \rceil$ bit locations, which divide the successive conditional ambiguity sets as close to half as possible. Set $J^l$ is defined as:
\begin{equation*}
J^l = \{j^l_k, 1 \le k \le \lceil \log \mu^l_{X_1,\ldots, X_N} \rceil \},
\end{equation*}
where $j^l_k$ is defined as:
\begin{equation}
j^l_k = \argmin_{i \in U|\{b^*(j^l_1), \ldots, b^*(j^l_{k-1}) \}} |N_i^0 - N_i^1|
\end{equation}
and $b^*(j^l_t) \in \{0,1\}, t = 1, \ldots, k-1$, is defined as follows:
\begin{equation}
\label{eqn:bitValue}
b^*(j^l_t) = \argmax\limits_{s = \{0,1\}} \widehat{\mu}_{{X_1,\ldots, X_N}|\{b^*(j^l_1), \ldots, b^*(j^l_{t-1}), b(j^l_t) = s\}}
\end{equation}
We summarize this formally as ``\textbf{Round-Parallel}'' protocol, given below.

\hspace{-1.0em}\hrulefill

\hspace{-0.5em}{\textbf{Protocol:} Round-Parallel}

\vspace{-0.2cm}\hspace{-1.0em}\hrulefill
\begin{codebox}
\li $l = 0$.
\li Let $S_{X_1,\ldots, X_N}^l = S_{X_1,\ldots, X_N}$
\li Let $V = \{1, \ldots, \sum_{i = 1}^{N}{\cal I}_{X_i}\}$
\li Let $U$ be the set of undefined bits in $V$, $U \subseteq V$, over all $\overline{x} \in S_{X_1,\ldots, X_N}^l$
\li \While ($\mu_{X_1,\ldots, X_N}^l > 1$)
\li \hspace{-0.45in}  $J^{l+1} = \phi$
\li \hspace{-0.45in}  \For ($k = 1, \ldots, \lceil \log \mu_{X_1,\ldots, X_N}^l \rceil$)
\li \hspace{-0.8in}    $j_k^{l+1} = \argmin\limits_{i \in U|\{b^*(j_1^{l+1}), \ldots, b^*(j_{k-1}^{l+1}) \}} |N_i^0 - N_i^1|$, where $b^*(j_t^{l+1})$, $t = 1, \ldots, k-1$
\zi \hspace{-0.8in}    is defined as in \eqref{eqn:bitValue}
\li \hspace{-0.8in}    Compute $U \subset V$, the set of undefined bits \label{parCom:undefBits}
     \End
\li \hspace{-0.45in}  The sink asks the informants corresponding to the bit-locations in $J^{l+1}$ \label{parCom:bitValue}
\zi \hspace{-0.45in}  to send the bit-values $b(j^{l+1}_k)$, $k = 1, \ldots, \lceil \log \mu_{X_1,\ldots, X_N}^l \rceil$
\li \hspace{-0.45in}  Compute $S_{X_1,\ldots, X_N}^{l+1} = \bigcap\limits_{k = 1}^{\lceil \log \mu_{X_1,\ldots, X_N}^l \rceil} \!\!\! S_{{X_1,\ldots, X_N}|b(j^{l+1}_k)}^l$
\li \hspace{-0.45in}  Compute $U \subset V$, the set of undefined bits
\li \hspace{-0.45in}  $l = l + 1$
    \End
\end{codebox}
\vspace{-0.2cm}\hrulefill

The worst-case behavior of \textbf{Round-Parallel} protocol can be analyzed by assuming the set of informant bits in Line~\ref{parCom:bitValue} is same as the set of their worst-case values, that is, $b(j^{l+1}_k) = b^*(j^{l+1}_k)$.

\textit{Upper bound on number of rounds}: Suppose that the data-gathering finishes in $k$ communication rounds and in every round, the informants send ${\cal I}_{X_1,\ldots, X_N}$ bits. Assume that in $i^\textrm{th}, 1 \le i\le k$, communication round, the size of the ambiguity set is reduced by $2^{(1-\epsilon_i)\lceil \log \mu_{X_1,\ldots, X_N} \rceil}, \epsilon_i < 1$. So, after $k$ rounds, we have $\frac{\mu_{X_1,\ldots, X_N}}{2^{\sum_{s=1}^k (1-\epsilon_s)\lceil \log \mu_{X_1,\ldots, X_N} \rceil}} = 1$.

Define $\epsilon = \max\{\epsilon_1, \ldots, \epsilon_k\}$. Assume that the size of the ambiguity set in every round is reduced by $2^{(1-\epsilon)\lceil \log \mu_{X_1,\ldots, X_N} \rceil}$. Assume that the data-gathering finishes now in $k^*$ rounds. It is obvious that $k \le k^*$. Now, the size of the ambiguity set after $k^*$ rounds satisfies $\frac{\mu_{X_1,\ldots, X_N}}{(2^{(1-\epsilon)\lceil \log \mu_{X_1,\ldots, X_N} \rceil})^{k^*}} = 1$. So,
\begin{equation*}
k^* = \bigg\lceil \frac{\log \mu_{X_1,\ldots, X_N}}{(1-\epsilon){\cal I}_{X_1,\ldots, X_N}}\bigg\rceil \le \bigg\lceil \frac{1}{(1-\epsilon)}\bigg\rceil
\end{equation*}
This implies that $k^* = 1, \mbox{ if } \epsilon \le 0$; $k^* \le 2, \mbox{ if } 0 < \epsilon, \epsilon \approx 0$; and $k^* \le \big\lceil \frac{1}{(1-\epsilon)}\big\rceil, \mbox{ if } \epsilon \approx 1$.

\textit{Upper bound on number of informant bits}: The total number of informant bits sent in a round is upper-bounded by ${\cal I}_{X_1,\ldots, X_N}$, so the total number of informant bits sent over all rounds $\sum_{l = 1}^{k} \lceil \log \mu^l_{X_1,\ldots, X_N} \rceil$ is upper-bounded by  $k^* {\cal I}_{X_1,\ldots, X_N}$.

\textit{Upper bound on number of sink bits}: The sink can address each of $N$ informants in $\lceil \log N \rceil$ bits. So, it addresses all informants in $N \lceil \log N \rceil$ bits. In $N$ more bits, it informs all informants whether those have to transmit anything in the current communication round or not. The sink asks the informant $i$ in $\lceil \log \log \mu_{X_i} \rceil$ bits, to send the bit-value corresponding to the bit-index $\lceil \log \log \mu_{X_i} \rceil$. So, the total number of bits that the sink sends over all rounds, is upper-bounded by $k^* (N \lceil \log N \rceil + N + \sum_{i \in J} \lceil \log \log \mu_{X_i} \rceil)$. In the case, when all the informants encode their information in $\lceil \log n \rceil$ bits each, then the total number of sink bits is upper bounded by $k^* (N \lceil \log N \rceil + N + {\cal I}_{X_1,\ldots, X_N} \lceil \log \log n \rceil)$.

\begin{pavikl}
\label{lemma:parComBoundsbSerCom}
The total number of informant bits under \textbf{Round-Parallel} protocol upper-bound the total number of informant bits under \textbf{Bit-Serial} protocol.
\end{pavikl}
\begin{IEEEproof}
In \textbf{Bit-Serial} protocol, the optimal bit-location (in the sense of dividing the resultant ambiguity set as close to half as possible) to be polled in a round is determined by actual values of the previously polled optimal bit-locations. However, in \textbf{Round-Parallel} protocol, in the $l^\textrm{th}$ round, $l \ge 1$, out of $\lceil \log \mu^l_{X_1,\ldots, X_N} \rceil$ bit-locations to be polled, all except the first bit-location to be polled are selected by assuming that the previously chosen bit-location assume their worst-case bit-values, as in \eqref{eqn:bitValue}. This implies that \textbf{Round-Parallel} protocol, while provisioning for the worst-case, over-estimates the total number of informant bits, compared to \textbf{Bit-Serial} protocol. Therefore the number of informant bits under \textbf{Round-Parallel} protocol upper-bound the total number of informant bits under \textbf{Bit-Serial} protocol, thus proving the lemma.
\end{IEEEproof}

\begin{pavikc}
\label{cor:parComBoundsbSerCom}
The performance of \textbf{Round-Parallel} is same as that of \textbf{Bit-Serial} on those elements of the support-set on which latter achieves its worst-case performance, in terms of total number of informant bits.
\end{pavikc}
\begin{IEEEproof}
For those members of the support-set on which \textbf{Bit-Serial} protocol performs the worst, \textbf{Round-Parallel} protocol while provisioning for the worst-case, precisely chooses the values of same bit-locations to be communicated as \textbf{Bit-Serial} protocol, thus achieving identical performance.
\end{IEEEproof}

All parallel protocols require more total number of informant bits, in the worst-case, compared to \textbf{Bit-Serial}. However, among all such round parallel protocols, \textbf{Round-Parallel} provides the best worst-case performance, as next lemma states.

\begin{pavikl}
\label{lemma:parComOptimal}
The \textbf{Round-Parallel} protocol is optimal round parallel communication protocol.
\end{pavikl}
\begin{IEEEproof}
We prove the theorem by considering its following implication. No round parallel protocol can do better than \textbf{Round-Parallel} protocol in the following sense: the total number of informant bits and the number of rounds it requires for a given support-set are no less than as required by \textbf{Round-Parallel} protocol for all elements of the given support-set.

Assume for the sake of contradiction that there is a round parallel communication protocol, let us call it \textbf{Protocol X}, that is better than \textbf{Round-Parallel} protocol. This implies that \textbf{Protocol X} achieves at least one of the following:
\begin{enumerate}
\item[Case 1:] \hspace{0.05in} Under \textbf{Protocol X} fewer informant bits are sent in fewer communication rounds compared to \textbf{Round-Parallel} protocol.
\item[Case 2:] \hspace{0.05in} Under \textbf{Protocol X} fewer informant bits are sent in same number of communication rounds compared to \textbf{Round-Parallel} protocol.
\item[Case 3:] \hspace{0.05in} Under \textbf{Protocol X} same number of informant bits are sent in fewer communication rounds compared to \textbf{Round-Parallel} protocol.
\end{enumerate}

Now, we prove that each of these three cases leads to a contradiction.

\textit{Case 1:} If \textbf{Protocol X} sends fewer informant bits than \textbf{Round-Parallel} protocol for all elements of the support-set, then this implies that even for the elements on which \textbf{Bit-Serial} protocol or \textbf{Round-Parallel} protocol (from Corollary~\ref{cor:parComBoundsbSerCom}) achieves its worst-case performance, in terms of total number of informant bits, \textbf{Protocol X} can achieve better performance. However, this contradicts the worst-case optimality of \textbf{Bit-Serial} protocol (from Lemma~\ref{lemma:bSerComOptimal}).

\textit{Case 2:} Applying same reasoning as in Case 1 to this case leads to a similar contradiction.

\textit{Case 3:} For a given support-set if \textbf{Round-Parallel} protocol finishes the data-gathering in $k, k > 1$, rounds in the worst-case, then we can always construct a round parallel communication protocol \textbf{Protocol X} that finishes the data-gathering in $k', 1 \le k' < k$, rounds with same number of informant bits as \textbf{Round-Parallel} protocol in the worst-case. However, any such protocol while provisioning for the worst-case, ends-up sending more informant bits and requires more rounds than \textbf{Round-Parallel} protocol for those elements of the support-set on which \textbf{Round-Parallel} protocol does not achieve worst-case optimal performance.

Each of these cases shows that there are always some elements of the support-set on which \textbf{Protocol X} performs worse than \textbf{Round-Parallel} protocol. Therefore, we prove that no round parallel communication protocol can do better than \textbf{Round-Parallel} protocol for all elements of the given support-set.
\end{IEEEproof}

\section{Role of Block-coding in the Worst-case Distributed Compression}
\label{sec:blkCoding4cmprsn}
We have, thus far, discussed the notion of worst-case compressibility in distributed source coding scenario when only a single instance of data-vector is available at the informants (\textit{oneshot compression}). However, the majority of results in classical Information Theory are derived in the limit of asymptotic block-lengths, though recently the role of non-asymptotic block-lengths has been investigated \cite{110polyanskiyPoorVerdu, 110costaLangbergBarros, 110polyanskiyPoorVerdu2}. These results firmly establish the effectiveness of block-coding in achieving the optimal average-case performance of various information-theoretic problems. In this section, we attempt to investigate the effectiveness of block-coding in realizing the optimal worst-case performance of asymmetric distributed source coding problem.

Formally, we are concerned with addressing the question whether solving the worst-case bit-compressibility problem over block-length $k, k > 1$, results in fewer informant bits and larger achievable rate-region than solving this problem $k$ times over single instance of data as in \eqref{eqn:bitCompressibility}. To aid in our subsequent analysis, we introduce some definitions.

Define $S^1_{X_1, \ldots, X_N} = S_{X_1, \ldots, X_N}$. Then, for $k > 1$, the $k^{\mathrm{th}}$-extension of support-set $S_{X_1, \ldots, X_N}$ is:
\begin{equation}
\label{eqn:snkextn}
S^k_{X_1, \ldots, X_N} = S^{k-1}_{X_1, \ldots, X_N} \times S_{X_1, \ldots, X_N}
\end{equation}
The $k^{\mathrm{th}}$-extension of data-vector $\overline{x}, \overline{x} \in S_{X_1, \ldots, X_N}$, is:
\begin{equation*}
\overline{x}^k = (x_1, \ldots, x_N)^k = (x^1_1 \cdots x^k_1, \ldots, x^1_N \cdots x^k_N) = (\overline{x}^k_1, \ldots, \overline{x}^k_N)
\end{equation*}
Then, the $k^{\mathrm{th}}$-extension of support-set $S_{X_i}, i \in {\mathcal N}$, is:
\begin{equation}
\label{eqn:s1kextn}
S^k_{X_i} = \{\overline{x}^k_i | \mbox{ for some } \overline{x}^k_{-i}, (\overline{x}^k_i, \overline{x}^k_{-i}) \in S^k_{X_1, \ldots, X_N} \}
\end{equation}
Note that $|S^k_{X_1, \ldots, X_N}| = \mu^k_{X_1, \ldots, X_N}$ and $|S^k_{X_i}| = \mu^k_{X_i}$.

\textbf{Problem Statement:} Consider a distributed information-gathering scenario where a sink collects the data from $N$ correlated informants. As in Section~\ref{sec:probSetting}, divide the sequence of events in this distributed data-compression problem in terms of \textit{data-generation epoch} and \textit{data-gathering epoch}. However, in this case, the data-generation epoch is repeated $k$ times, $k > 1$, before each data-gathering epoch where the sink learns each of $k$ strings of each informant. Rest of the details of the problem setting are same as in Section~\ref{sec:probSetting} and we do not repeat those here.

A sample $\overline{x} = (x_1, \ldots, x_N)$ is drawn from the discrete and finite support-set $S_{X_1, \ldots, X_N}$ over $N$ binary strings. The strings of $\overline{x}$ are revealed to the informants, with string $x_i$ being given to the $i^{\mathrm{th}}$ informant. This process is repeated $k$ times, $k \ge 1$, resulting in each informant accumulating $k$ instances of its data. At the end of this data-generation epoch, the sink begins data-gathering to learn each of $k$ strings of each informant. We are interested in minimizing the total number of informant bits required, in the worst-case, to enable the sink to \textit{losslessly} ($P_e = 0$) learn each of the $k$ instances of $\overline{x}$ revealed to the informants in the previous $k$ data-generation epochs.

Consider an alternative problem formulation in terms of a new problem-encoding scheme that also facilitates the design and analysis of optimal solutions in our setting.

\textbf{Alternate Problem Statement:} Assume that $k$-extended data-vector $\overline{x}^k = (\overline{x}_1^k, \ldots, \overline{x}_N^k), \overline{x}_i^k = (x^1_i, \ldots, x^k_i)$, is drawn from discrete and finite $k$-extended support-set $S^k_{X_1, \ldots, X_N}$ over $N$ binary-strings. The strings of $\overline{x}^k$ are revealed, unbeknownst to the sink, to the informants with the string $\overline{x}_i^k$ given to the $i^{\mathrm{th}}, i \in {\mathcal N}$, informant.

The sinks knows that one of the strings from $S^k_{X_1, \ldots, X_N}$ is drawn and its different components are given to different informants. The objective of data-gathering is to enable the sink to learn the identity of this string by communicating with different informants.

In our asymmetric communication scenario, the sink knows support-set $S^k_{X_1, \ldots, X_N}$. We assume that each informant $i, i \in {\mathcal N}$, knows its $k$-extended support-set $S_{X_i}^k$. The sink can uniquely describe every $\overline{x}^k, \overline{x}^k \in S^k_{X_1, \ldots, X_N}$, in $\lceil k \log \mu_{X_1, \ldots, X_N}\rceil$ bits. However, for the same reasons as in Section~\ref{sec:probSetting} to efficiently query the informants for the purpose of data-gathering, the sink can also uniquely encode every $\overline{x}^k$ in terms of set $B_{\overline{x}^k}$ of $\sum_{i \in {\mathcal N}} \lceil k \log \mu_{X_i} \rceil$ bits by concatenating $\lceil k \log \mu_{X_i} \rceil$ bits long representation of each $\overline{x}_i^k, i \in {\mathcal N}$.

\begin{table*}[!t]
\centering
\caption{Example of Problem Encoding Scheme for DSC with Block-coding where column (a) refers to Support set $S_{X_1, X_2}$, (b) refers to elements of $S_{X_1, X_2}^2$ (2-extended support-set), (c) refers to 2-Block Data of Informants, (d) refers to Binary Representation, and (e) refers to Concatenated Representation}
\vspace{0.4cm}
\begin{tabular}{ c | c | c | c | c }
\hline
(a) & (b) & (c) 	& (d) 			&   (e) \\\hline
\multirow{11}{*}{\includegraphics[width=1.5in]{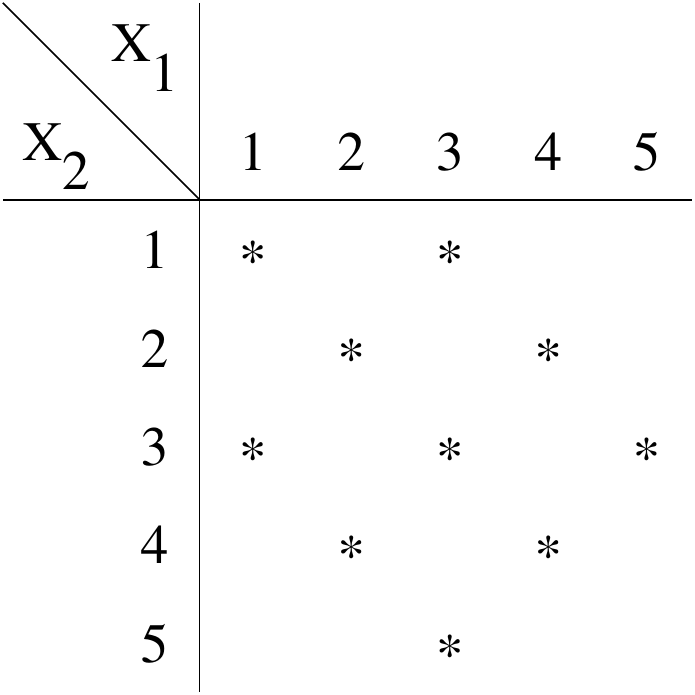}}&(1,1) (1,1)	&	(1,1)  (1,1) & 	(00000) (00000) & 0000000000 \\
&(1,1) (1,3)	&	(1,1)  (1,3) & 	(00000) (00010) & 0000000010 \\
&(1,1) (2,2)	&	(1,2)  (1,2) & 	(00001) (00001) & 0000100001 \\
&(1,1) (2,4)	&	(1,2)  (1,4) & 	(00001) (00011) & 0000100011 \\
&(1,1) (3,1)	&	(1,3)  (1,1) & 	(00010) (00000) & 0001000000 \\
&(1,1) (3,3)	&	(1,3)  (1,3) & 	(00010) (00010) & 0001000010 \\
&(1,1) (3,5)	&	(1,3)  (1,5) & 	(00010) (00100) & 0001000100 \\
&(1,1) (4,2)	&	(1,4)  (1,2) & 	(00011) (00001) & 0001100001 \\
&(1,1) (4,4)	&	(1,4)  (1,4) & 	(00011) (00011) & 0001100011 \\
&(1,1) (5,3)	&	(1,5)  (1,3) & 	(00100) (00010) & 0010000010 \\
&$\vdots$	     &	$\vdots$     & 	$\vdots$        & $\vdots$   \\\hline
\end{tabular}
\label{table:kBlkProbEncoding}
\end{table*}

\textbf{Example 2:} We illustrate the proposed problem-encoding scheme for alternative problem statement above in Table~\ref{table:kBlkProbEncoding}. Consider the support-set $S_{X_1, X_2}$ of two jointly distributed random variables $X_1$ and $X_2$ defined over alphabet ${\mathcal X} = \{1, 2, 3, 4, 5\}$, with $|S_{X_1, X_2}| = 10$, as given in the first column. The elements of $2$-extension of this support-set $S_{X_1, X_2}^2, |S_{X_1, X_2}^2| = 100$, are listed in the second column, however for the sake of brevity, we list only 10 of these elements. Recall that each element of $S_{X_1, X_2}^2$ is the concatenation of two samples of $\overline{x} = \{x_1, x_2\}$. In the third column, we list the corresponding $2$-extended data-block at each informant. The sink and each informant can agree \textit{a priori} on some deterministic binary-encoding of $2$-extension of informant's data. In the fourth column, we give one such encoding and the fifth column lists the concatenation of binary-encoding of data-blocks at each informant. \endproof

With this encoding scheme, the worst-case bit-compressibility problem with block-length $k$ is to identify the smallest subset of bit-locations of size $C_B^k$ in the concatenated bit-representation of $\overline{x}^k$, whose values the sink must know to unambiguously learn $\overline{x}^k$ revealed to the informants. That is,
\begin{equation}
\label{eqn:kblk_bCompressibility}
C_B^k = \max_{\overline{x}^k \in S^k_{X_1, \ldots, X_N}} \min_{b_{\overline{x}^k} \subseteq B_{\overline{x}^k}} |b_{\overline{x}^k}| \mbox{ such that } S^k_{{X_1, \ldots, X_N}|b_{\overline{x}^k}} = \{\overline{x}^k\}
\end{equation}
where $S^k_{{X_1, \ldots, X_N}|b_{\overline{x}^k}}$ is the $k$-extended conditional ambiguity set when the subset $b_{\overline{x}^k}, b_{\overline{x}^k} \subseteq B_{\overline{x}^k}$, is known at the sink. In the next subsection, we discuss the solution of \eqref{eqn:kblk_bCompressibility}.

\subsection{Worst-case Bit-Compressibility with Block-coding}
\label{subsection:kBlk_bCompressibility}
From the previous discussion, it is easy to observe that $C_B^k$ satisfies: $\lceil k \log \mu_{X_1, \ldots, X_N}\rceil \le C_B^k \le \sum_{i = 1}^N \lceil k \log \mu_{X_i} \rceil$. Therefore, for asymptotic block-lengths, $k \rightarrow \infty$, $k$-block bit-compressibility per block $\frac{C_B^k}{k}$ satisfies:
\begin{equation*}
\log \mu_{X_1, \ldots, X_N} \stackrel{(a)}{\le} \lim_{k \rightarrow \infty} \frac{C_B^k}{k} \stackrel{(b)}{\le} \sum_{i = 1}^N \log \mu_{X_i},
\end{equation*}
where (a) follows from $\lim_{k \rightarrow \infty} \frac{\lceil{kx}\rceil}{k} = x$ and (b) follows from 
\begin{align*}
\frac{1}{k}\sum_{i = 1}^N \lceil k \log \mu_{X_i} \rceil &\le \sum_{i = 1}^N \log \mu_{X_i} + \frac{N}{k} \\
\lim_{k \rightarrow \infty} \frac{1}{k}\sum_{i = 1}^N \lceil k \log \mu_{X_i} \rceil &\le \sum_{i = 1}^N \log \mu_{X_i}
\end{align*}
as $\lim_{k \rightarrow \infty} \frac{N}{k} = 0$ for any finite and fixed $N$.

Recall from Section~\ref{sec:probSetting} that the worst-case bit-compressibility for oneshot compression satisfies: $\lceil \log \mu_{X_1, \ldots, X_N}\rceil \le C_B \le \sum_{i = 1}^N \lceil \log \mu_{X_i} \rceil$. This implies that compared to oneshot computation, the block-coding improves the lower-bound on bit-compressibility by no more than one bit and the corresponding upper-bound is reduced by at most one bit per informant. This leads us to conclude that for the worst-case distributed compression problem in our setting, the block-coding offers \textit{almost} no gain with respect to oneshot compression. Therefore, the oneshot compression is \textit{almost} optimal with respect to solving the worst-case DSC problem in our setting.

To formally prove that $C_B - \frac{C_B^k}{k} \le 1$ for all $k > 1$, a communication protocol to optimally solve the problem in \eqref{eqn:kblk_bCompressibility} can be designed as a $k, k > 1$, block generalization of \textbf{Bit-Serial} protocol introduced in Section~\ref{sec:bCompressibility4cmprsn} for solving the worst-case DSC problem for $k = 1$ (\textit{oneshot compression}). Replacing the various support-sets and ambiguities in \textbf{Bit-Serial} protocol by their $k$-extended equivalents and using the problem-encoding scheme proposed above in this section, results in the desired protocol, called ``\textbf{$k$-extended Bit-Serial}'' protocol, given below.

\hspace{-1.0em}\hrulefill

\hspace{-0.5em}{\textbf{Protocol:} $k$-extended Bit-Serial}

\vspace{-0.2cm}\hspace{-1.0em}\hrulefill
\begin{codebox}
\li $l = 0$
\li Let $S_{X_1,\ldots, X_N}^{(k,l)} = S_{X_1,\ldots, X_N}^k$
\li Let $V = \{1, \ldots, \sum_{i = 1}^{N} \lceil k \log \mu_{X_i} \rceil\}$: index set of all bit-locations in $B_{\overline{x}^k}$
\li Let $U$ be the index set of undefined bits in $V$, $U \subseteq V$
\li \While ($\mu_{X_1,\ldots, X_N}^{(k,l)} > 1$)
\li \hspace{-0.45in} $J^{l+1} = \argmin_{i \in U} |N_i^0 - N_i^1|$
\li \hspace{-0.45in}  If $|J^{l+1}| > 1$, then choose uniformly at random the bit-location $j^{l+1}, j^{l+1} \in J^{l+1}$
\li \hspace{-0.45in}  The sink asks the informant corresponding to bit-location $j^{l+1}$ to send bit-value $b(j^{l+1})$
\li \hspace{-0.45in}  Set $S_{X_1,\ldots, X_N}^{(k,l+1)} = S_{{X_1,\ldots, X_N}|b(j^{l+1})}^{(k,l)}$
\li \hspace{-0.45in}  Compute $U \subset V$, the set of undefined bits
\li \hspace{-0.45in}  $l = l + 1$
    \End
\end{codebox}
\vspace{-0.2cm}\hrulefill

The proof of optimality of \textbf{$k$-extended Bit-Serial} protocol for the worst-case $k$-block distributed compression is obtained by following the same reasoning that was used to prove the optimality of \textbf{Bit-Serial} protocol in Section~\ref{sec:bCompressibility4cmprsn}. Therefore, we omit it here.

\begin{figure}[!t]
\centering
\includegraphics[width=4.0in]{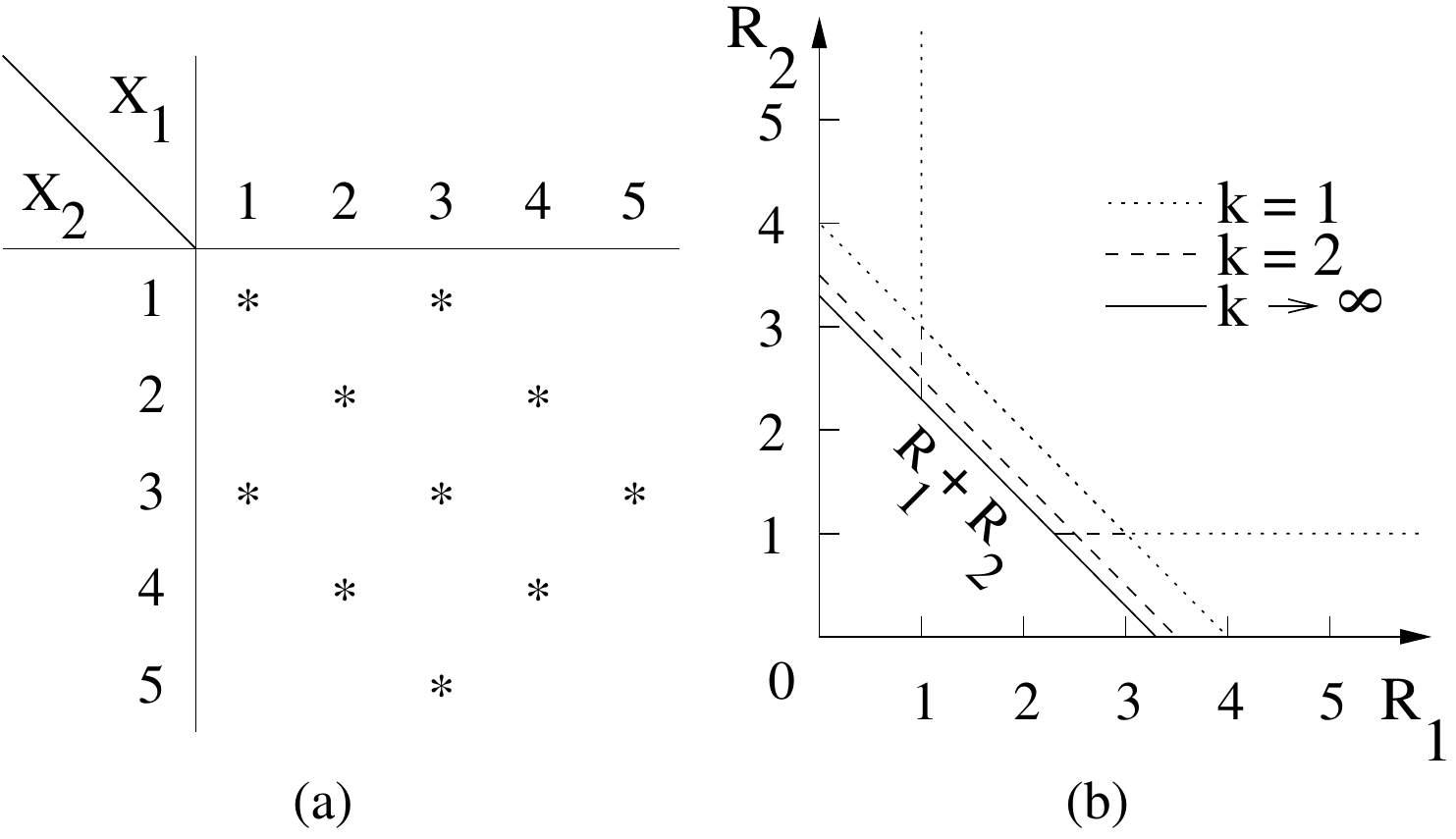}
\vspace{-0.1in}
\caption{(a) Support-set $S_{X_1, X_2}$ with $\mu_{X_1, X_2} = 10$ (b) corresponding worst-case achievable rate-regions for data-block length $k = 1, 2$ and $k \rightarrow \infty$.}
\label{fig:dist6blkCoding}
\vspace{-0.1in}
\end{figure}

\begin{figure}[!t]
\centering
\includegraphics[width=4.0in]{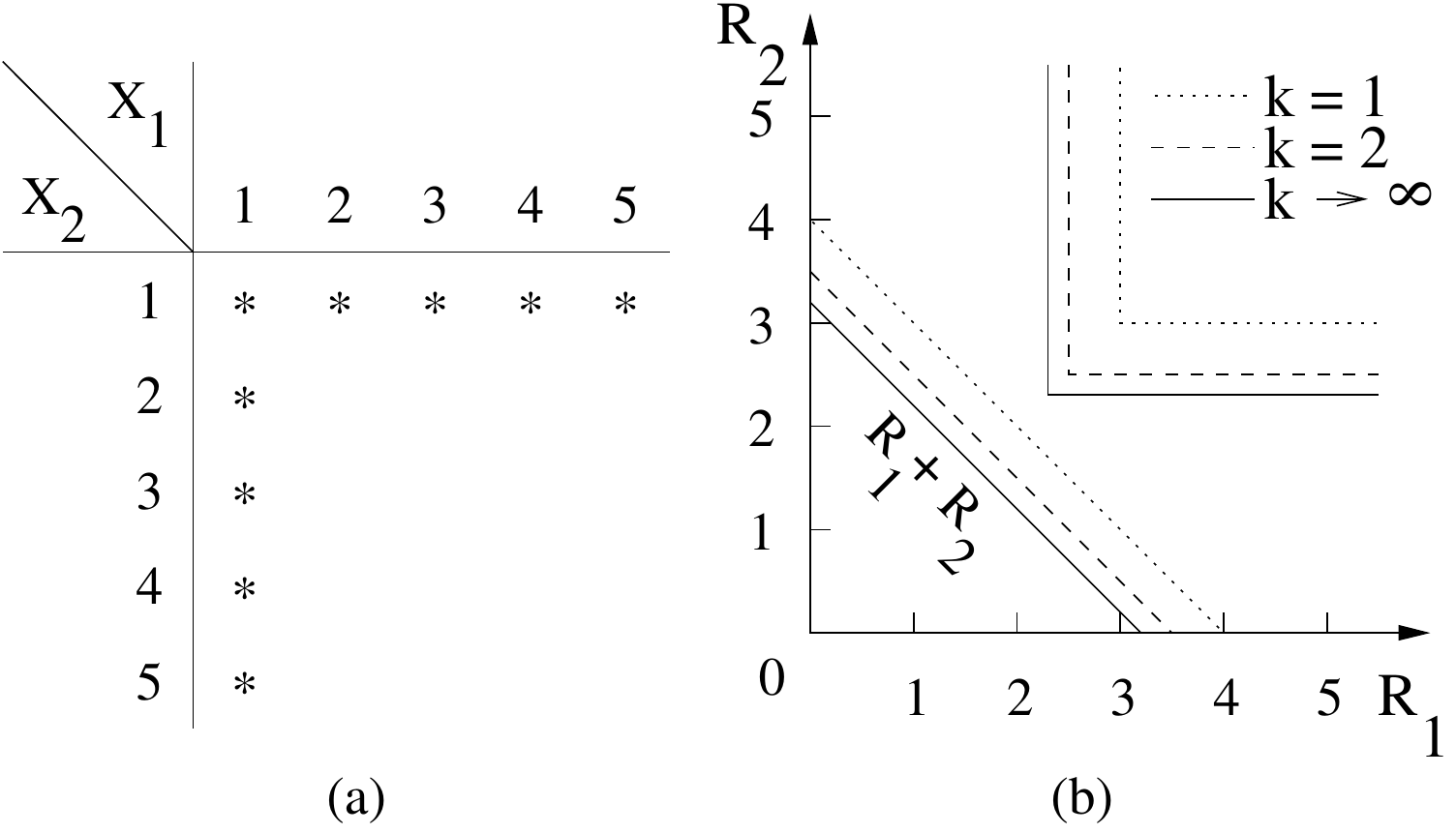}
\vspace{-0.1in}
\caption{(a) Support-set $S_{X_1, X_2}$  with $\mu_{X_1, X_2} = 9$ (b) corresponding worst-case achievable rate-regions for data-block length $k = 1, 2$ and $k \rightarrow \infty$.}
\label{fig:dist3blkCoding}
\vspace{-0.1in}
\end{figure}

Using \textbf{$k$-extended Bit-Serial} protocol, in Figures~\ref{fig:dist6blkCoding}-\ref{fig:dist3blkCoding}, we plot the worst-case achievable rate-regions for two different support-sets of two correlated informants with block-length $k = 1, 2$, and $k \rightarrow \infty$. These figures demonstrate the limiting behaviour of sum and individual information rates as the function of block-length for two particular distributions and establish the \textit{almost} optimality of solving the worst-case distributed compression problem with only a single instance of informant data-vector.

Next, we compute the achievable rate-region for the worst-case DSC problem with block-coding in terms of the achievable rate-region for the worst-case distributed compression problem with single instance of informant data-vector. The following lemma states our result.

\begin{pavikl}
\label{lemma:kBlk-rateRegion}
Given the set of achievable rate-vectors for the worst-case distributed compression problem for the oneshot compression, as in Corollary~\ref{cor:NrateRegion}, the set of achievable rate-vectors for the worst-case distributed compression problem with $k, k > 1$, block-coding is defined by
\begin{equation*}
R^{k}(S) \ge \frac{\lceil k \log(2^{M_{R(S)} - 1} + 1) \rceil}{k}
\end{equation*}
for all $S \subseteq \mathcal{N}$, where $R^{k}(S) = \sum_{i \in S} R_i^k$
\end{pavikl}
\begin{IEEEproof}
Consider constraint $R_1 \ge M_{R_1}$. The size of the set that can be described in $M_{R_1}$ bits from Informant 1 lies between $2^{M_{R(S)} - 1} + 1$ and $2^{M_{R(S)}}$. Therefore, the size of the $k$-extension of this set lies between ${(2^{M_{R(S)} - 1} + 1)}^k$ and $2^{kM_{R(S)}}$. This implies that the minimum number of bits per block required from Informant 1 to describe the $k$-extended set is at least $\frac{k \lceil \log(2^{M_{R_1)} - 1} + 1) \rceil}{k}$.

Identical argument holds for proving other constraints in the statement of the lemma for all other subsets of $\mathcal{N}$. Combining all the proofs together, proves the lemma.
\end{IEEEproof}

\section{Conclusions and Future Work}
\label{sec:conclusions}
We consider classical problem of Distributed Source Coding in Information Theory. We propose a new canonical scheme to construct and address different variants of this problem.

The classical distributed source coding (DSC) problem in Information Theory finds a natural application in addressing the data-gathering problem in wireless sensor networks, where sensor data is often assumed to be correlated. However, existing approaches to address distributed source coding problem cannot be employed directly to solve the data-gathering problem in wireless sensor networks. In this paper, therefore, we propose a variant of distributed source coding problem that works with single instance of sensor data-vector to reduce the latency of data-gathering, employs interactive communication to reduce expenditure of communication and computation resources at the nodes, and does not require sensor nodes to have the complete knowledge of the entire network. Further, to perform the worst-case information-theoretic analysis of certain problems in wireless sensor networks, we propose the notion of \textit{information ambiguity}, prove that it is a valid information measure, and derive its various properties.

We provide optimum and constructive solutions of the proposed variant of the distributed source coding problem in two communication scenarios in terms of two respective protocols and prove that unlike the average-case performance of distributed source coding problems, the worst-case performance of such problems is not enhanced by employing block-coding and the optimal worst-case performance can be achieved just with a single instance of source data-vector.

We have also proposed a system-architecture to implement our work in actual data-gathering wireless sensor networks to enhance their lifetime. However, the details of such extensions of our work are beyond the scope of this paper and are discussed in one of our forthcoming submissions.

\textit{Future Work:} We are currently working towards generalizing classical Information Theory in some newer directions to carry out its more meaningful applications to various other problems that cannot be addressed in the existing framework. In particular, we are working towards extending the results in the current paper to distributed compression problems where the informants do not communicate directly with the sink but do so via some intermediate nodes. The operations that the intermediate nodes can perform, depending on their computational capabilities, on informants' data determine the maximum compression that can be achieved. We are also attempting to generalize the notion of information ambiguity to discrete and infinite, and continuous support-sets as such generalizations have interesting implications in some problems in distributed inference and learning. We plan to continue to work on such problems.

We have striven for a systems-level understanding of the problem of communication under very general conditions and a systems-level solution to the problem. Nowhere did the problem formulation nor the proposed solution rely on details and specifics of the agents that make up the system. If we change the nature of the probability distributions or the objective functions of the resulting optimization problem, the system can represent diverse models of the real-world situations. Therefore, our solution techniques, approaches, and heuristics are independent of the specific application domains. Our continuing goal will be to apply our framework to understand communication in complex systems like intra-cellular communication and neural information processing systems among others.  

We recognize that our framework is in its infancy. However our approach suggests a systematic and principled framework for generating generalized information-like measures that are fine-tuned to aid in the task of optimal design and analysis of systems with communicating agents. Depending on the underlying objective functions and constraints of the optimization problem, we expect a zoo of information-like measures will arise. As readers may have noticed, the proofs and results in this work uses tools from several disciplines. As we generalize our work, we expect to find intricate and deep connections between these disciplines and hope to pursue it concurrently in the future.

\end{document}